\newcommand{\rb}{{\bf r}}
\newcommand{\ub}{{\bf u}}

\newcommand{\Eb}{{\bf E}}

\newcommand{\Hb}{{\bf H}}
\newcommand{\Ib}{{\bf I}}

\newcommand{\vb}{{\bf v}}

\newcommand{\nhat}{\hat{\bf{n}}}
\newcommand{\uhat}{\hat{\bf{u}}}
\newcommand{\xhat}{\hat{\bf{x}}}
\newcommand{\yhat}{\hat{\bf{y}}}
\newcommand{\Hhat}{\hat{\bf{H}}}

\newcommand{\omegab}{\boldsymbol{\omega}}

\newcommand{\nablab}{\mbox{\boldmath $\nabla$\unboldmath}} 

\newcommand{\Gammab}{\mbox{\boldmath $\Gamma$\unboldmath}}

\newcommand{\beq}{\begin{equation}}

\newcommand{\eeq}{\end{equation}}

\newcommand{\bea}{\begin{eqnarray}}

\newcommand{\eea}{\end{eqnarray}}

\PassOptionsToPackage{sort&compress}{NJDnatbib}

\documentclass[ASNA,twocolumn]{USG} 
\usepackage{anyfontsize} %

\usepackage{colortbl}   
\usepackage{xcolor}     

\usepackage{steinmetz}

\usepackage{orcidlink}

\articletype{RESEARCH ARTICLE}%

\begin{document}

\title{Shape Asymmetry Turns Curved Flows and Orienting Fields into Versatile Tools for
Microswimmer Sorting, Trapping, and Phase Control}


\transtitle{Manuscript Title}

\subtranstitle{trans-subtitle}

\author[1]{Derek C. Gomes}[][\orcidlink{0009-0006-6318-2624}]

\author[1]{Tapan C. Adhyapak}[][\orcidlink{0000-0002-8251-2880}]

\authormark{TAYLOR \textsc{et al.}}
\titlemark{PLEASE INSERT YOUR ARTICLE TITLE HERE}

\address[1]{\orgdiv{Department of Physics,}\orgname{ Indian Institute of Science Education and
Research (IISER) Tirupati, }%
\orgaddress{\state{Andhra Pradesh 517619,}\country{India}}}

\presentaddress{This is sample for present address text this is sample for present address text.}

\keywords{Microswimmers | Active Particles | Shape-Asymmetry | Microfluidics  | Bacteria | Low Reynolds number swimmers | Microswimmer sorting | Microswimmer trapping | Magnetotaxis, Phototaxis, Electrotaxis | }

\transkeywords{Keyword1 | Keyword2 | Keyword3 | Keyword4 | Keyword5}

\abstract[ABSTRACT]{Controlling fluidic transport of active particles requires
understanding how particle shape and motility modify their response to imposed
flows and fields. In this work, we show how the coupling of microswimmer shape
asymmetry to imposed flows, and its interplay with an orienting field, give
rise to distinct microswimmer dynamics that can be leveraged for particle
sorting by shape and motility in channel flows. We identify the steady-state
heights and orientations accessible to swimmers in the channel and determine
how they vary with the strength and orientation of the applied field, using
complementary simulations and theory. We further construct the dynamical phase
spaces of four swimmer classes, distinguished by shape asymmetry and motility,
and identify their full set of attractors and basins of attraction. Connecting
these structures to swimmer trajectories, we predict the relative abundances of
different population states and demonstrate their control by external
field-flow parameters. From these analyses, we uncover several previously
overlooked hydrodynamic phenomena, including \textit{horizontal layering} of
the channel into regions with distinct orientational dynamics, and demonstrate
applications such as a \textit{hydrodynamic trap} that uses imposed flows to
pin swimmers with a selected propulsion speed at a prescribed location within
the channel.}

 \transabstract[transABSTRACT]{This is a generic template designed for use by multiple journals,
which includes several options for customization. Please
refer the author guidelines and author LaTeX manuscript preparation document for the journal to which you
are submitting in order to confirm that your manuscript will
comply with the journal’s requirements. Please replace this
text with your abstract. This is sample abstract text just for the template display purpose.}

\abbr{5-FU, 5-fluorouracil; CFD, computational fluid dynamics; CH, channel; EFS, event-free survival; GBM, glioblastoma multiforme; OS, overall survival; PFS, progression-free
survival; SD, standard deviation.}



\maketitle

\section{Introduction}
\label{sec:intro}

The controlled transport of microswimmers in imposed flows, guided by
externally applied orienting fields, is central to many biological processes
\cite{Bees2020Bioconvection, DurhamKesslerStockerScience2009,
KesslerNature1985} and emerging biomedical technologies
\cite{Felfoul2016TumorDelivery, Faivre2018Biohybrid, Li2017MicroNanorobots}.
It also advances the broader goal of programmable active matter
\cite{Yang2025ActiveMattProgramming,
Beppu2024MagneticallyControlledBacterialTurbulence,
Han2021ProgrammableChiralStates, Brady2023ConfinedActMattFields}.  However,
existing studies have largely overlooked particle shape asymmetry, a feature
that our recent work demonstrates can fundamentally reshape the dynamics of
self-propelled microswimmers in imposed flows \cite{GomesAdhyapakPRL2025}. Such
effects could enable dynamic sorting of microswimmers by shape and motility,
yet this capability remains limited in scope despite its potential impact
across healthcare, biotechnology, and active-particle technologies
\cite{Janca2010RoleShapeBacteriaSeparation,
Misko2025MotilityDependentSelectiveTransport,
Schuster2003IsolationMotileSpermatozoa, Berg1991SelectionMotileNonchemotactic}.
Furthermore, the potential of control fields to harness shape-asymmetry effects
remains largely underexplored. Our recent framework provides a first step in
this direction, showing that shape asymmetry generates previously overlooked
polar alignments in curved flows that can compete with externally applied
orienting fields, giving rise to rich microswimmer dynamics
\cite{GomesAdhyapak_Field-Particle2026}. However, the consequences of this
flow--field competition, and its potential for sorting, trapping, and
controlling microswimmer dynamics, have not been systematically explored.

\smallskip 

\textbf{In this paper}, we investigate the dynamical phases of head--tail
shape-asymmetric (HT-A) microswimmers, contrasting them with those of
head--tail shape-symmetric (HT-S) microswimmers and non-motile HT-A particles.
We consider imposed flows and orienting fields at arbitrary relative
orientations, and show how their competition can be harnessed to sort these
particle classes and control their dynamical states. We provide a comprehensive
map of HT-A dynamics in terms of experimentally tunable parameters, extending
our previous findings \cite{GomesAdhyapakPRL2025,
GomesAdhyapak_Field-Particle2026} toward practical applications. In addition,
we present the first phase-space maps of non-motile HT-A particles and HT-S
microswimmers over variable field orientations, revealing distinct dynamical
states and their \textit{basins of attraction} that provide a basis for sorting
and dynamical-state control across these particle classes. We further examine
the origin and nature of the dynamical attractors in these systems and, in the
process, identify \emph{hydrodynamic trapping} as an additional application of
shape-mediated flow--field competition.

\smallskip

The imposed flow considered here is a pressure-driven (Poiseuille) microchannel flow, which is common in both laboratory and natural settings. Such flows are widely used to control microswimmer transport because of their experimental accessibility and versatility \cite{SquiresQuake2005, StoneStroockAjdari2004}, as well as their relevance to naturally occurring environments \cite{Stocker2015MicrobesInFlow}. Moreover, the simple channel and flow geometry provides a versatile platform for probing the fundamental physics of transport, confinement, and swimmer--flow interactions in active suspensions \cite{Saintillan2015TransportDiluteActiveSuspension}.

\smallskip

By contrast, externally applied orienting fields, when implemented in these microchannel flows, enable control of microswimmer dynamics and transport through a range of physical cues. Many biological microswimmers naturally orient in response to magnetic \cite{BlakemoreScience1975}, gravitational \cite{PedleyHillKessler1988}, or optical \cite{Dervaux2022Phototaxis} fields, while synthetic swimmers can be designed to respond to an equally broad range of external stimuli \cite{Ybert2025MacroShear, ranabir_PRL2024, Bechinger2016Phototaxis, Yu2019Thermophoresis, Thutupalli_drop_squirm}. Together, imposed flows and orienting fields constitute a central paradigm for controlling microswimmer transport in biological and microfluidic environments \cite{Faivre2024MagBacReview}.

\smallskip 

\smallskip 

Most studies of microswimmer dynamics in field--flow settings focus almost exclusively on HT-S swimmers, or treat HT-A swimmers within an HT-S approximation \cite{Faivre2024MagBacReview, Guasto2012Review, salima_prl, PedleyKesslerReview1992}. This idealization overlooks the pronounced head--tail shape asymmetry of many suspended active particles, including artificial \cite{Guan2023MagTadpole, Shelley2019AuPtRods,Nelson2016Acoustic} and bioengineered microswimmers \cite{AubryACS2020}, as well as prototypical living microswimmers such as \textit{E. coli} and \textit{B. subtilis}, which are widely used as model systems in active-matter research \cite{Clement2026ScaleFree, Wu2017WeakSync, Aranson2022BacterialAM, tapanPRE2015}. Our recent studies \cite{GomesAdhyapakPRL2025, GomesAdhyapak_Field-Particle2026} established that shape-asymmetry-driven dynamics can become pronounced within experimentally accessible parameter regimes, whereas neglecting shape asymmetry can lead to oversimplified descriptions. How shape asymmetry can be harnessed through externally tunable flow--field parameters to control microswimmer dynamics and enable dynamic sorting has remained unexplored and is a central focus of this study.

\medskip

\textbf{Our main findings are:}

\begin{itemize}

\item \textit{Non-motile HT-S particles.} HT-S particles follow streamlines. In the presence of a field, particles near the channel midplane adopt height-dependent fixed orientations, whereas those farther from the midplane tumble.

\item \textit{HT-S microswimmers.} Motile HT-S particles exhibit two trajectory families: wall-focused and collimated. Motility makes these trajectories quadrant selective: for $\phi_0$ in the first and fourth quadrants, all trajectories focus at the walls, whereas in the second and third quadrants they collimate at a unique, field-selected height. Motility greatly reduces the wall-focusing length and amplifies oscillations about the mean trajectory. For collimated swimmers, these oscillations decay, yielding a steady upstream swimmer orientation.

\item \textit{Non-motile HT-A particles.} By contrast, the imposed flow generates shape-asymmetry-enabled HT-A polar torques and a polar force, both absent for HT-S particles. At zero field, as a result of these effects, particles tumble uniformly about their hydrodynamic centres, independent of their initial channel height, thus driving periodic cross-stream oscillations with no net transverse drift. A nonzero field competes with the HT-A polar torques, producing particle-orientation-dependent rotation rates and net transverse drift. The resulting trajectories either collimate at one \textit{or more} tunable heights or focus at the channel walls over tunable \emph{focal lengths}; both trajectory classes occur across all field-direction quadrants, unlike the quadrant-selective HT-S dynamics. At a fixed field direction $\phi_0$, the selected trajectory is determined by the initial particle height and orientation, with different initial conditions falling within distinct basins of attraction for the different dynamical states.

\item \textit{HT-A microswimmers.} Shape-asymmetry-enabled, flow-induced polar torques compete with the applied field, while the corresponding polar drift competes with microswimmer motility, producing dynamical states absent for motile HT-S swimmers and non-motile HT-A particles. In addition to wall-focused and sharply off-center collimated trajectories across all field-direction quadrants, stable oscillatory trajectories---a new trajectory class---arise in the second and third quadrants, forming an extended collimated beam in the channel centre. Remarkably, motile HT-A swimmers reach the walls over markedly shorter focusing lengths than motile HT-S swimmers, for comparable field strengths. At sufficiently strong fields and swimming speeds, the HT-A polar torques and force are masked, and the dynamical classes collapse onto those of motile HT-S swimmers.

\item \textit{Phase-space mapping and applications.} We map the phase space of each particle class across field strengths and directions, and use nonlinear dynamical analyses to identify the accessible attractors and explain their origins. The resulting framework enables (i) sorting motile swimmers by aspect ratio and shape symmetry (HT-S versus HT-A), as well as sorting motile and non-motile HT-A particles; (ii) hydrodynamic trapping that selectively captures motile swimmers with a target self-propulsion speed; and (iii) control of dynamical states by tuning field strength, field direction, imposed-flow speed, and channel width.

\end{itemize}

\subsection{Paper Organization}\label{subsec:paper_organisation}

The paper is organised as follows. In \textbf{Sect. \ref{sec:MicroswimmerDyn}}, we introduce the dynamics of microswimmers with general axisymmetric shapes in arbitrary fluid flows and applied fields, and use this framework to derive the equations of motion for two broad classes of HT-S and HT-A swimmers, respectively, without loss of generality. In \textbf{Sect. \ref{sec:Results}}, we systematically examine the dynamics of non-motile and motile particles, with and without head--tail asymmetry, identifying their dynamical attractors and elucidating the roles of motility, shape asymmetry, flow, and applied fields. We then use these dynamics to develop application-oriented predictions for hydrodynamic trapping, particle sorting, and phase control. Finally, in \textbf{Sect. \ref{sec:conclusion}}, we summarize the principal findings, discuss their broader implications, and outline directions for future work.

\smallskip

\section{Particles with General Axisymmetric Shapes in Imposed Flows and
Orienting Fields}
\label{sec:MicroswimmerDyn}

Below, we first set up the equations of motion for motile axisymmetric particles with arbitrary HT-A shapes in generalized imposed flows. The equations for HT-S microswimmers and for non-motile HT-A and HT-S particles can be constructed from these equations by taking appropriate limits, which we specify in the relevant sections.

\subsection{Equations of Motion (EOM)}
\label{sec:EOM}

Microswimmers propel themselves by exerting thrust on the surrounding fluid and are advected and rotated by the imposed flow field $\ub(\rb)$. External fields, denoted by $\Hb$, can additionally orient the swimmer. The resulting forces and torques acting on the swimmer determine its translational and angular velocities. We formulate the equations of motion about the swimmer's \textit{hydrodynamic center} $\rb_h$, the center of rotation determined by the distribution of forces exerted by the swimmer on the surrounding fluid. The translational velocity of $\rb_h$ and the angular velocity of the swimmer orientation $\nhat$ are therefore given by
\begin{eqnarray}
\dot{\rb}_h&=&v^{\rm sp}\nhat+\vb_h^{\rm flow}, \label{eq:vh} \\
\dot{\nhat}&=&\left(\omegab_h^{\rm flow}+\omegab_h^{\rm field}\right)\times\nhat. \label{eq:omegah}
\end{eqnarray}
Here, $v^{\rm sp}$ is the self-propulsion speed along $\nhat$, which we take to be constant.

\smallskip

The above equations apply to microswimmers with an intrinsic self-propulsion direction $\nhat$, irrespective of their shape symmetry. Shape dependence enters through the freely suspended
translational and angular velocities, $\vb_h^{\rm flow}$ and $\omegab_h^{\rm flow}$, induced by the surrounding flow $\ub(\rb)$. For an HT-S shape, the dynamics are invariant under $\nhat\to-\nhat$, and $\rb_h$ coincides with the geometric center, whereas an HT-A shape breaks this symmetry. For the general HT-A case, following Ref.~\cite{GomesAdhyapak_Field-Particle2026}, the freely suspended velocities are
\begin{eqnarray}    
\vb^{\rm{flow}}_h   &=& \ub_h 
                        + \boldsymbol{\Gamma}_1\!\cdot\! \left(\nhat\!\cdot\!\nablab_h\right)\ub_h
                        + \boldsymbol{\Gamma}_2\!\cdot\! \nabla_h^2 \ub_h 
                        + \boldsymbol{\Gamma}_3\!\cdot\! \left(\nhat\!\cdot\!\nablab_h\right)^2\ub_h, \label{eq:vh_flow}\\
\omegab_h^{\rm{flow}} &=& \frac{1}{2} \nablab_h \times \ub_h 
                          + \boldsymbol{\Gamma}_4\!\cdot\!\left(\nhat\!\cdot\!\nablab_h\right)\nablab_h\times \ub_h \nonumber\\
                          &&+ \, \nhat \times 
                            \left[  \kappa_1 \Eb_h \!\cdot\! \nhat
                                  + \kappa_2 \nabla^2_h \ub_h
                                  + \kappa_3 \left( \nhat \!\cdot\! \nablab_h \right)^2 \ub_h\right].
\label{eq:omegah_flow}
\end{eqnarray}
  Here, $\ub_h=\ub(\rb_h)$, $\nablab_h\equiv\nablab_{\rb_h}$, and
  $\Eb_h=[\nablab_h\ub_h+(\nablab_h\ub_h)^T]/2$ is the strain-rate tensor at
  $\rb_h$. The tensors
  $\boldsymbol{\Gamma}_i=\Gamma_i^{\parallel}\nhat\nhat+\Gamma_i^\perp
  (\Ib-\nhat\nhat)$ and the constants $\kappa_i$ depend on the particle's shape.

\smallskip

Eqs.~\eqref{eq:vh_flow} and \eqref{eq:omegah_flow} contain all symmetry-allowed terms up to second order in the spatial gradients of the flow $\ub(\rb)$, a valid truncation when $\ub(\rb)$ varies slowly over the particle length scale. The $\Gammab_1$, $\Gammab_4$, and $\kappa_{2,3}$ terms are not invariant under $\nhat\to-\nhat$ and constitute the only HT-A contributions; the remaining terms are HT-S and correspond to those appearing in the classical Faxén--Jeffery equations \cite{kim_karilla}. In particular, the $\Gammab_1$ term represents a flow-gradient-driven \textbf{HT-A force}, whereas the $\Gammab_4$ and $\kappa_{2,3}$ terms represent flow-curvature-driven \textbf{HT-A torques}---that is, force and torque contributions that are distinct under $\nhat\to-\nhat$.

\smallskip

\begin{figure}
\centerline{\includegraphics[width=\columnwidth]{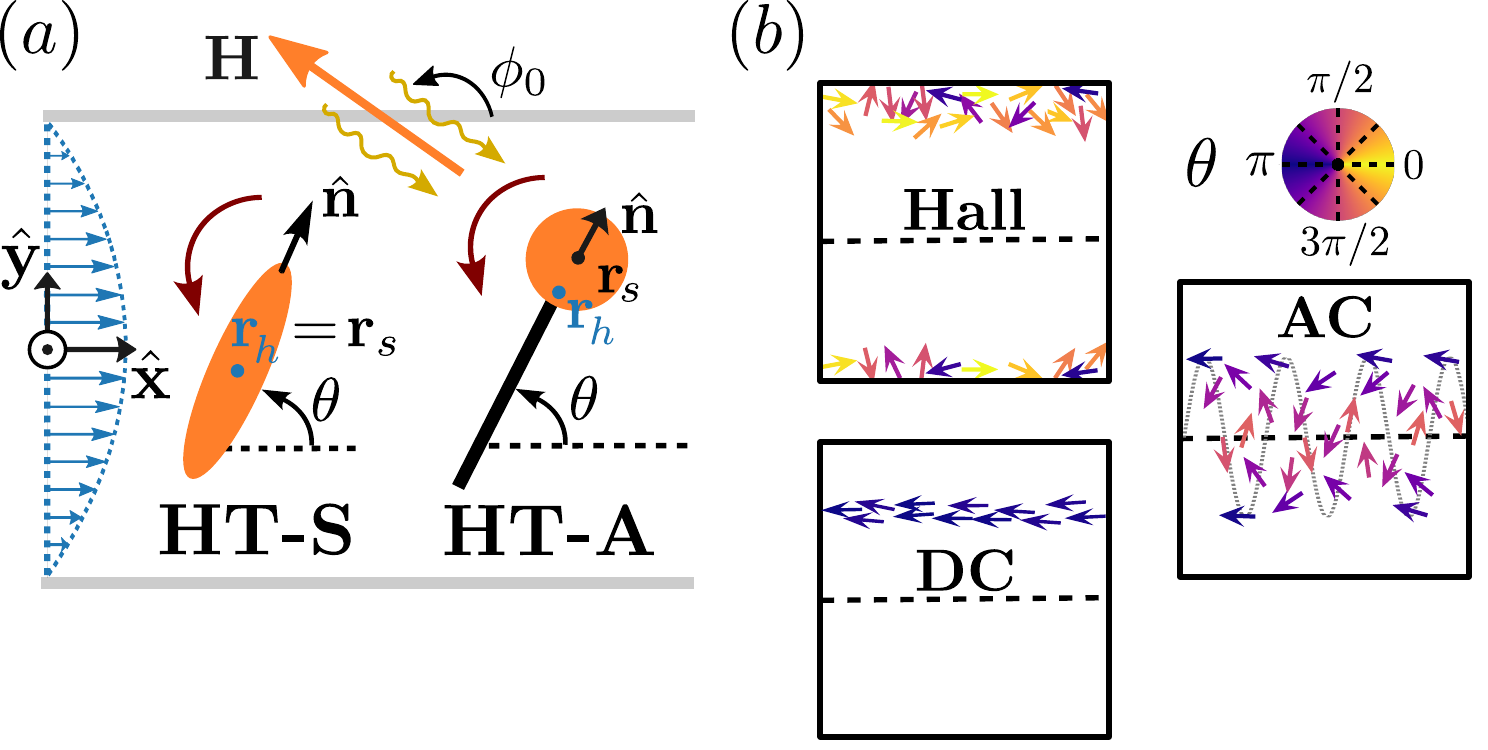}}

\caption{ (a) Schematic of HT-S and HT-A microswimmers in channel flow under an
applied orienting field $\Hb$ of strength $\mathcal{H}$, incident at an angle
$\phi_0$ to the flow direction. (b) Schematic of the distinct steady states
accessible to HT-S and HT-A swimmer populations: Hall, AC, and DC states. Arrow colors
indicate swimmer orientation (see color wheel).}

\label{fig1}

\end{figure}
\textbf{External orienting field and its effects.} In all examples of external orienting fields discussed in Sec.~\ref{sec:intro}, the field exerts an effective aligning torque proportional to $\nhat\times\Hb$ on the particle. Denoting the proportionality constant by an effective particle \emph{dipole moment} $\mu^{\rm eff}$, we write the field-induced torque as $\mu^{\rm eff}\nhat\times\Hb$. The corresponding angular velocity is
\begin{eqnarray}
\omegab_h^{\rm field}&=&\nhat\times\mathcal{H}\Hhat ,
\label{eq:omegaField}
\end{eqnarray}
where $\mathcal{H}=\mu^{\rm eff}|\Hb|/\gamma_R$ and $\gamma_R$ is the particle's rotational friction coefficient.

\smallskip

As shown in Ref.~\cite{GomesAdhyapak_Field-Particle2026}, the flow-induced HT-A torques associated with the $\kappa_{2,3}$ terms in Eq.~\eqref{eq:omegah_flow} tend to align swimmers along or against the local flow direction $\ub_h$ and compete directly with the field-induced torque associated with $\omegab_h^{\rm field}$, which aligns swimmers with $\Hb$. Similarly, the flow-induced HT-A force associated with the $\Gammab_1$ term in Eq.~\eqref{eq:vh_flow} competes with the self-propulsion force associated with $v^{\rm sp}\nhat$. In the Results, we analyse the consequences of these field--torque and activity--force competitions as functions of field strength and direction.

\smallskip

\textbf{Dynamics in channel flow.} Henceforth, for the imposed flow $\ub(\rb)$, we consider Poiseuille flow, introduced earlier as a ubiquitous flow in natural and technological systems. Furthermore, it is one of the simplest flow profiles with finite curvature, which is essential for the HT-A couplings to contribute to the swimmer orientation dynamics, as seen from Eq.~\eqref{eq:omegah_flow}. We therefore set $\ub(\rb)=v_f[1-(\rb\cdot\yhat)^2/R^2]\xhat$, corresponding to a microchannel of half-width $R$ whose midplane coincides with the $y=0$ plane [Fig.~\ref{fig1}(a)]. In the high-P\'eclet-number regime (large $v_f$), we focus on the dominant shear-plane dynamics \cite{GomesAdhyapakPRL2025,shear-induced}, with $\rb_h\equiv(x,y)$ and $\nhat\equiv(\cos\theta,\sin\theta)$, where $\theta$ is measured from the flow direction $\uhat\equiv\xhat$ [Fig.~\ref{fig1}(a)]. Similarly, we define the field-direction angle $\phi_0$ through $\Hhat\equiv(\cos\phi_0,\sin\phi_0)$.

\smallskip

\subsection{HT-S Model: EOM and Exact Coefficients}

For an HT-S particle, $\vb_h^{\rm flow}$ and $\omegab_h^{\rm flow}$ are obtained from Eqs.~\eqref{eq:vh_flow} and \eqref{eq:omegah_flow} by setting $\Gammab_{1,4}=0$ and $\kappa_{2,3}=0$. To obtain the exact coefficients for the remaining terms, we represent, without loss of generality, the HT-S shape by an ellipsoid with semi-major and semi-minor axes $c$ and $b$, respectively. Expanding the standard HT-S expressions for $\vb_h^{\rm flow}$ and $\omegab_h^{\rm flow}$ for an ellipsoid \cite{kim_karilla} to second order in flow gradients then yields: $\Gammab_2=b^2\Ib/6$, $\Gammab_3=(c^2-b^2)\Ib/6$, and $\kappa_1=(c^2-b^2)/(c^2+b^2)$. The resulting $\vb_h^{\rm flow}$ and $\omegab_h^{\rm flow}$ for the HT-S particles, evaluated for channel flow $\ub(\rb)$ and substituted into Eqs.~\eqref{eq:vh} and \eqref{eq:omegah}, yield the shear-plane dynamics for HT-S particles as
\begin{eqnarray}
\dot{y}&=&     v^{\rm{sp}}\sin\theta, 
    \label{eq:ydotHTS_poiseuille}\\
\dot{\theta} &=& \frac{v_f}{R^2} \left[ \left(1-G\right)  
               + 2 G \sin^2{\theta} \right]y  -\mathcal{H}\sin\left(\theta - \phi_0 \right),
      \label{eq:thetadotHTS_poiseuille}
\end{eqnarray}
with an $x$-dynamics
\begin{eqnarray}
    \dot{x}=
v^{\rm sp}\cos\theta + \left[1 
                        + \frac{b^2}{6}\nabla_s^2 
                        + \frac{(c^2-b^2)}{6} \left(\nhat\!\cdot\!\nablab_s\right)^2\right]\ub_s\!\cdot\!\xhat,\label{eq:xdotHTS_poiseuille}
\end{eqnarray} 
decoupled from the $y,\theta$ dynamics. Thus,
$(y,\theta)$ form the phase space of the system. Note that, in the angular
equation, $G=\kappa_1=(c^2-b^2)/(c^2+b^2)$ recovers the standard Jeffery
coefficient.

\subsection{HT-A Sphere--Rod Model: EOM and Exact Coefficients}

For axisymmetric HT-A swimmers, all terms in Eqs.~\eqref{eq:vh_flow} and
\eqref{eq:omegah_flow} contribute in general. In particular, the HT-A
coefficients $\Gamma_{1,4}$ and $\kappa_{2,3}$ appearing in these equations
depend on the degree of head--tail shape asymmetry (i.e., the relative sizes
of the head and tail) and on the position of the hydrodynamic center relative
to the head and tail. To obtain their exact expressions, we consider a
sphere--rod model for the HT-A shape, shown schematically in Fig.~\ref{fig1}(a).
The model captures the essential HT-A features of a broad class of biological
\cite{omori_upward_sperm_swim, Najafi2019MultipleBundles, Deng2020Pseudomonas, tapanPRE2017}
and synthetic
\cite{Guan2023MagTadpole, Shelley2019AuPtRods,Nelson2016Acoustic}
flagellated microswimmers. The swimmer cell body is modelled as a sphere of
radius $a$, while its flagellar bundle is modelled as a slender rod of length
$\ell_f$ and cross-section $0.01a$. The corresponding coefficients $\Gammab_i$
and $\kappa_i$ are derived for this model in terms of $a$ and $\ell_f$ and
provided in Ref.~\cite{GomesAdhyapak_Field-Particle2026}.The particle orientation $\nhat$ is directed from the rod tip towards the sphere centre $\rb_s$, also governing its propulsion direction $v^{\rm sp}\nhat$. We note that our model is sufficiently broad to not restrict the generality of the findings listed above.

\smallskip 

Now, following Ref.~\cite{GomesAdhyapak_Field-Particle2026}, we recast the general
field--flow dynamics w.r.t. the cell-body centre $\rb_s$, which is
experimentally the most accessible point to track on such flagellated
swimmers. Because of the swimmer's HT-A shape, $\rb_s$ does not, in general,
coincide with $\rb_h$. However, since the swimmer is rigid, the translational
velocities of the two reference points $\rb_s$ and $\rb_h$ are related by
rigid-body kinematics, while the angular velocity remains unchanged about
either point. Thus, Eqs.~\eqref{eq:vh}--\eqref{eq:omegah}, now expressed in terms
of $\rb_s$, are given by:
\begin{eqnarray}
    \dot{\rb}_s&=&\dot{\rb}_h + \left(\omegab_h^{\rm flow}+\omegab_h^{\rm field}\right)\times r_{\rm{sh}}\nhat, \label{eq:vs} \\
    \dot{\nhat}&=& \left(\omegab_h^{\rm flow}+\omegab_h^{\rm field}\right)\times\nhat \label{eq:omegas}
\end{eqnarray}
where $r_{\rm sh}=|\rb_s-\rb_h|$, with the velocities $\vb^{\rm flow}_h$ and
$\omegab^{\rm flow}_h$ evaluated from the flow and its spatial derivatives
at $\rb_h$. To express the flow and its gradients, and hence the swimmer
dynamics, entirely w.r.t. the cell-body centre $\rb_s$, we use Taylor
expansions of these quantities from $\rb_h$ to $\rb_s$
\cite{GomesAdhyapak_Field-Particle2026}. Thus, Eqs.~\eqref{eq:vh_flow} and
\eqref{eq:omegah_flow}, with the flow and its derivatives now evaluated
w.r.t. $\rb_s$, have the modified coefficients (now denoted by primes):
$\boldsymbol{\Gamma}_1^{\prime} =\boldsymbol{\Gamma}_1-r_{\rm sh}\Ib$,
$\boldsymbol{\Gamma}_3^{\prime} =\boldsymbol{\Gamma}_3-r_{\rm sh}\boldsymbol{\Gamma}_1
+r_{\rm sh}^2\Ib/2$, $\boldsymbol{\Gamma}_4^{\prime}
=\boldsymbol{\Gamma}_4-r_{\rm sh}\Ib/2
+r_{\rm sh}\kappa_1(\Ib-\nhat\nhat)/2$, and
$\kappa_3^{\prime} =\kappa_3-\kappa_1 r_{\rm sh}$, with all other coefficients
remaining unchanged. For the channel-flow setup for $\ub(\rb)$ described above,
with $(x,y)$ now denoting $\rb_s$, substituting Eqs.~\eqref{eq:vh_flow} and
\eqref{eq:omegah_flow} with the primed coefficients into
Eqs.~\eqref{eq:vs} and \eqref{eq:omegas}, and with
$\nhat\equiv(\cos\theta,\sin\theta)$ as in the HT-S case, yields the
required shear-plane EOMs for HT-A swimmers:
\begin{eqnarray}
\dot{y}      &=&     v^{\rm{sp}}\sin\theta + r_{\rm{sh}}\dot{\theta}\cos\theta
                  + \dot{y}^{\rm{HTS}}_{\rm{flow}} + \dot{y}^{\rm{HTA}}_{\rm{flow}},
                    \label{eq:ydotHTA_poiseuille}\\
\dot{\theta} &=& \frac{v_f}{R^2} \left[ \left(1-G\right)  y
               + 2 G \sin^2{\theta} y + G_1 \sin{\theta}
               + G_2\sin^3{\theta}\right] \nonumber \\
                 && -\mathcal{H}\sin\left(\theta - \phi_0 \right),
\label{eq:thetadotHTA_poiseuille}
\end{eqnarray}
with an $x$-dynamics
\begin{eqnarray}
    \dot{x}=
v^{\rm sp}\cos\theta-r_{\rm sh}\dot{\theta}\sin\theta + \dot{x}^{\rm{HTS}}_{\rm{flow}}+\dot{x}^{\rm{HTA}}_{\rm{flow}},
\label{eq:xdotHTA_poiseuille}
\end{eqnarray} 
decoupled from the $y,\theta$ dynamics.  Thus, as in the HT-S case, $(y,\theta)$ form the phase space of the HT-A dynamics.

\smallskip

In the above, $G=\kappa_1$ recovers the standard Jeffery coefficient for the
HT-A sphere--rod model, while $G_1=2\kappa_2$ and
$G_2=2(\kappa_3-r_{\rm sh}\kappa_1)$ characterize its HT-A curvature
couplings. As mentioned above, the forms of $\kappa_{1-3}$ are provided in
Ref.~\cite{GomesAdhyapak_Field-Particle2026}. The flow contributions to the
translational velocities have the expressions:
\begin{eqnarray}
    \dot{y}^{\rm{HTS}}_{\rm{flow}}&=&\mathcal{A}_y\frac{v_f}{2R^2}
         \left[\frac{a^2}{3}
        -\left(a^2+a\ell_f+\frac{\ell_f^2}{2}\right)
         \sin^2\theta\right]\sin{2\theta}, \label{expr:ydotHTS_flow}\\
    \dot{y}^{\rm{HTA}}_{\rm{flow}}&=&\mathcal{A}_y\frac{v_f}{R^2}
        \left(a+\frac{\ell_f}{2}\right) y
       \sin\theta\sin{2\theta}, \label{expr:ydotHTA_flow}\\
    \dot{x}^{\rm{HTS}}_{\rm{flow}}&=&\left[1+\frac{a^2}{6}\nabla^2_s\right]\ub(\rb_s)\cdot \xhat + \left(\mathcal{A}_x-\mathcal{A}_y\sin^2{\theta}\right)  \nonumber \\
    &&\times\frac{v_f}{2R^2}
         \left[\frac{a^2}{3}
        -\left(a^2+a\ell_f+\frac{\ell_f^2}{2}\right)
         \sin^2\theta\right]\sin{2\theta}, \label{expr:xdotHTS}\\
    \dot{x}^{\rm{HTA}}_{\rm{flow}}&=& \left(\mathcal{A}_x-\mathcal{A}_y\sin^2{\theta}\right) \frac{v_f}{R^2}
        \left(a+\frac{\ell_f}{2}\right) y
       \sin\theta\sin{2\theta}. \label{expr:xdotHTA}
\end{eqnarray}
Here, $\mathcal{A}_{x,y}$ are swimmer-anisotropy prefactors, whose expressions,
together with the swimmer parameters $\Gammab_i$ and $\kappa_i$, are given in
Ref.~\cite{GomesAdhyapak_Field-Particle2026}. Notably, the HT-S contributions,
$\dot{y}^{\rm{HTS}}_{\rm{flow}}$ and $\dot{x}^{\rm{HTS}}_{\rm{flow}}$,
ultimately arise from the $\Gammab_{2,3}$ terms, while the HT-A contributions
arise from the $\Gammab_1$ term in Eq.~\eqref{eq:vh_flow}.

\subsection{Integration Method and Parameters}

The dynamical equations
\eqref{eq:ydotHTS_poiseuille}--\eqref{eq:xdotHTS_poiseuille} and
\eqref{eq:ydotHTA_poiseuille}--\eqref{eq:xdotHTA_poiseuille} are now fully
specified in terms of known swimmer, flow, and field parameters for HT-A and HT-S
swimmers, respectively. The corresponding trajectories
$\{x(t),y(t),\theta(t)\}$ are obtained by integrating the equations using
Euler's method with a time step of $dt=5\times10^{-5}\,\text{s}$. We use a
channel with flow strength $v_f=500\,\mu\text{m}/\text{s}$ and half-width
$R=50\,\mu\text{m}$. For the HT-A sphere--rod shape, we take $a=1\,\mu\text{m}$
and $\ell_f=5\,\mu\text{m}$. Using a rod cross-section $0.01 a$, consistent with typical
flagellar thickness and an ambient fluid viscosity equal to that of room
temperature water yields swimmer anisotropy parameters
$(\mathcal{A}_x,\mathcal{A}_y)=(0.538, -0.127)$
\cite{GomesAdhyapak_Field-Particle2026}. The resulting parameters are
$G=0.902$, satisfying $0<G<1$, $G_1=0.057\,\mu\text{m}>0$, and
$G_2=-4.577\,\mu\text{m}<0$; accounting for active flows places
$\rb_h$ at a distance $r_{\rm sh}=0.75a$ from $\rb_s$
\cite{GomesAdhyapakPRL2025}. For the HT-S ellipsoidal shape, $b=1\,\mu\text{m}$
and $c=b\sqrt{(1+G)/(1-G)}$, giving the same Jeffery coefficient
$G=0.902$ as for the HT-A case. For both HT-S and HT-A motile swimmers,
$v^{\rm sp}=50\,\mu\text{m}/\text{s}$, while $v^{\rm sp}=0\,\mu\text{m}/\text{s}$
for the corresponding non-motile cases.

\smallskip
\section{Results}
\label{sec:Results}

\begin{figure}

\centerline{\includegraphics[width=\columnwidth]{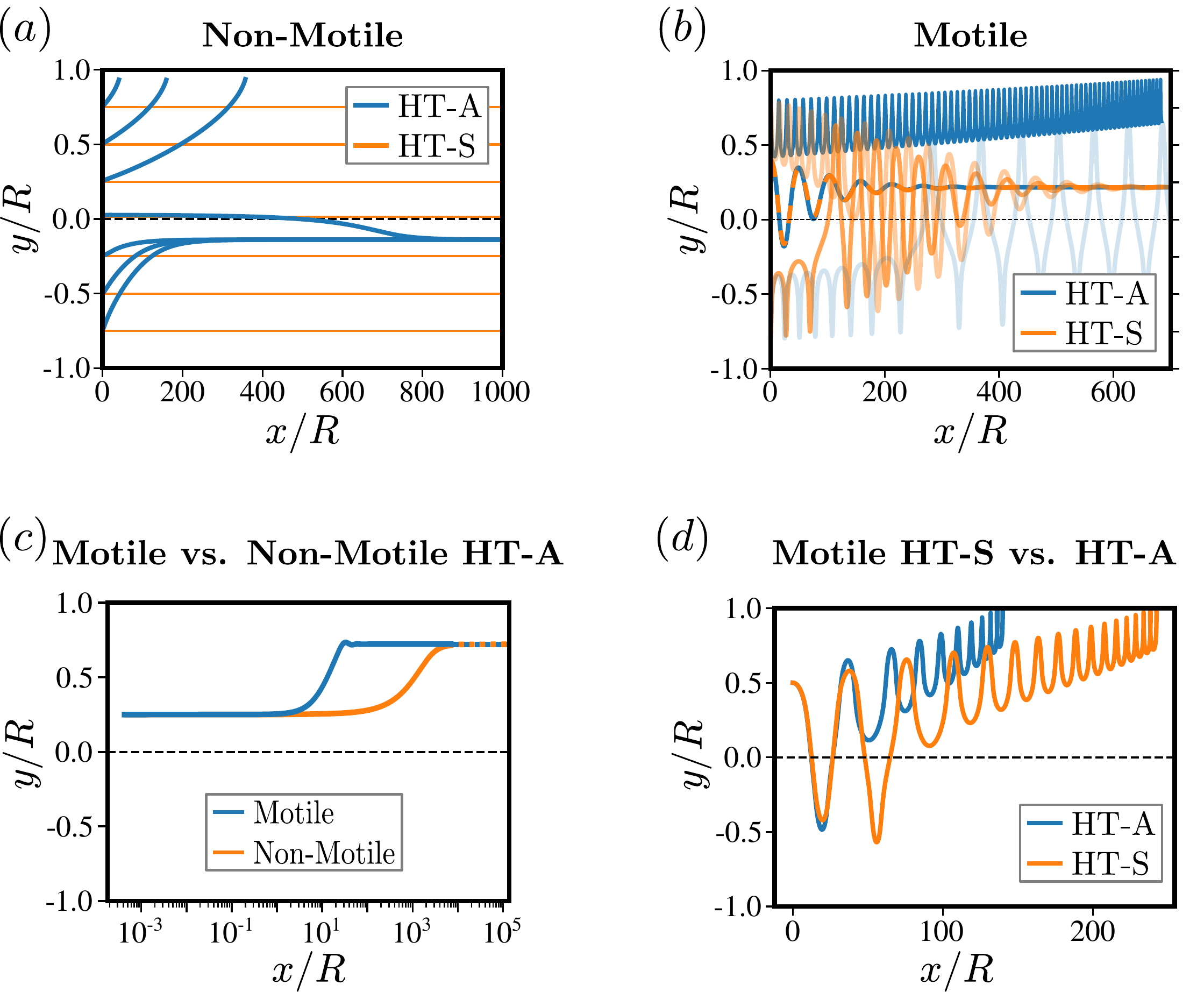}}

\caption{\textbf{Top row, trajectory overview:} HT-A (blue) and HT-S (orange) trajectories for (a) non-motile and (b) motile microswimmers, released from identical initial heights $y_i$ with $\theta_i=\pi$. Here, $\phi_0=3\pi/4$, with $\mathcal{H}=5.0\,\rm{s}^{-1}$ in (a) and $\mathcal{H}=0.3\,\rm{s}^{-1}$ in (b). \textbf{Bottom row, distinct focusing rates:} (c) Motile and non-motile HT-A swimmers, with $\mathcal{H}=1.0\,\rm{s}^{-1}$, $\phi_0=3\pi/4$, showing distinct off-midplane focusing rates; (d) motile HT-S and HT-A swimmers, with $\mathcal{H}=0.5\,\rm{s}^{-1}$, $\phi_0=\pi/4$, showing distinct wall-focusing rates.}

\label{fig2}
\end{figure}

The field--torque and activity--force competitions introduced in
Sec.~\ref{sec:EOM} were predicted and analysed for motile HT-A particles in
Ref.~\cite{GomesAdhyapak_Field-Particle2026}. Here, we extend this analysis to
all four particle classes summarized above. We systematically identify and
characterize the dynamical attractors accessible across field strengths and
directions, elucidate their origins, and use them to formulate
application-oriented predictions.

\smallskip

Figure~\ref{fig2} contrasts representative trajectories for the four
particle classes. Non-motile HT-S particles exhibit the familiar
streamline-following trajectories, whereas head--tail shape asymmetry alone
produces cross-stream motion that can focus particles either at the walls or
at a target plane displaced from the channel midplane
[Fig.~\ref{fig2}(a)]. Motility further enriches the dynamics: oscillations
about the mean trajectory are strongly amplified, while wall-focused and
off-midplane collimated states persist, and new stable oscillatory trajectories
emerge to form an extended collimated beam in the channel centre
[Fig.~\ref{fig2}(b)]. These dynamics enable shape-based sorting and
dynamical-state control, including accelerated off-midplane focusing of motile
HT-A particles relative to non-motile HT-A particles [Fig.~\ref{fig2}(c)] and
faster wall focusing of motile HT-A particles than motile HT-S particles
[Fig.~\ref{fig2}(d)]. Below, we analyse each particle class in detail.

\subsection{Non-motile HT-S Particles}

\begin{figure}

\centerline{\includegraphics[width=\columnwidth]{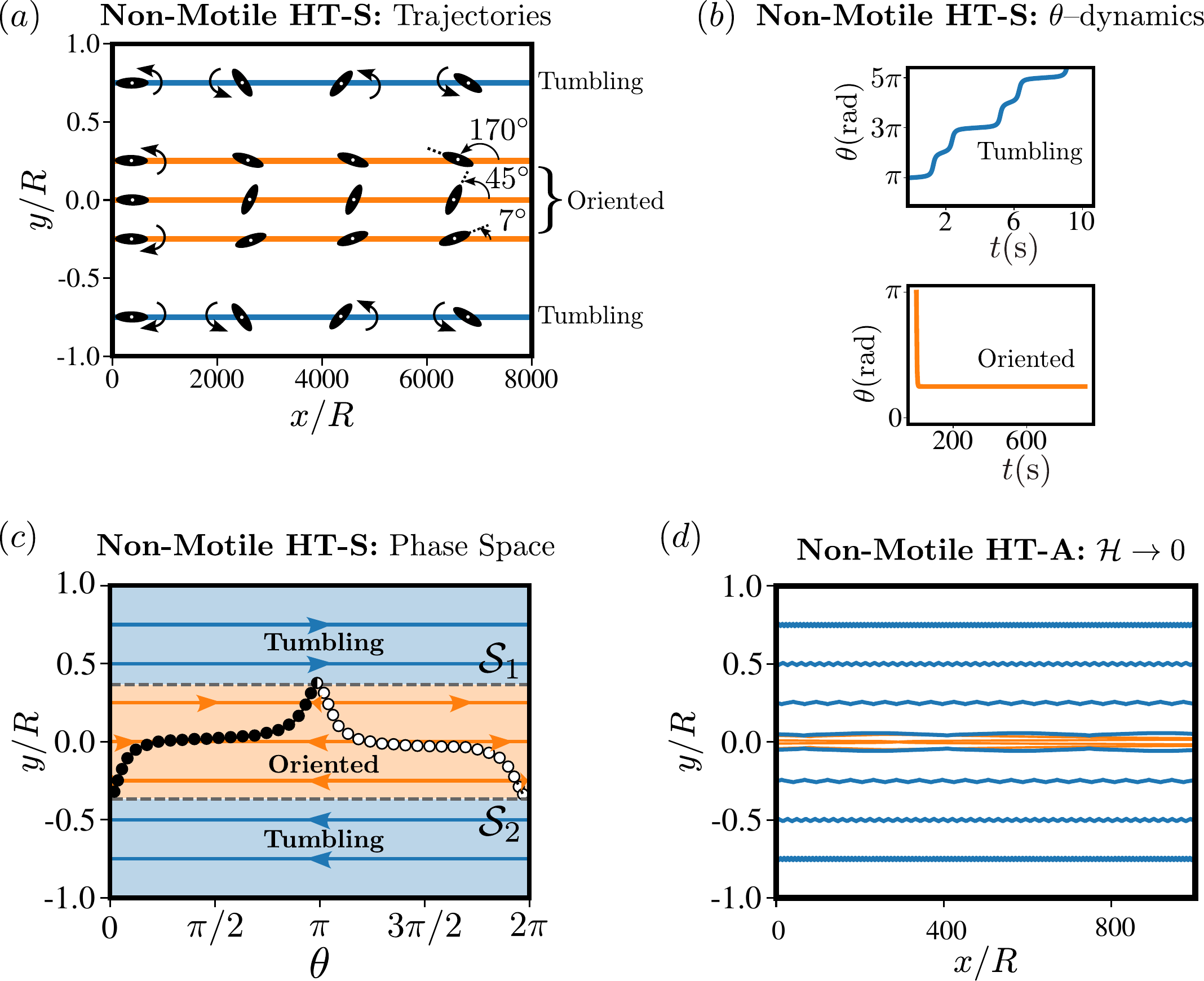}}

\caption{\textbf{Non-motile HT-S particles.} 
(a) Trajectories for particles released from different $y_i$ with $\theta_i=\pi$, for $\mathcal{H}=0.5,\rm{s}^{-1}$ and $\phi_0=\pi/4$. Orange and blue curves correspond to the \textit{Oriented} and \textit{Tumbling Regions}, respectively. (b) $\theta$ versus $t$ for $y_i=0.5R$ in the \textit{Tumbling Region} (left) and $y_i=0$ in the \textit{Oriented Region} (right). (c) The $y$--$\theta$ phase space, showing stable (filled) and unstable (open) fixed points in the \textit{Oriented Region} and separatrices $\mathcal{S}_1$ and $\mathcal{S}_2$ bounding the \textit{Tumbling Region}. (d) \textbf{Non-motile HT-A particles, $\boldsymbol{\mathcal{H}} \boldsymbol{\to 0}$:} blue trajectories oscillate without net cross-streamline drift, while orange trajectories drift only slightly towards nearby streamlines.}

\label{fig3}
\end{figure}

\textbf{Trajectory Classes and Dynamical Phase Space.}
Non-motile HT-S particles may be either passive or active without a preferred
swimming direction (shakers). As shown in Fig.~\ref{fig3}(a), non-motile
HT-S swimmers in field and flow simply follow streamlines. Remarkably, this is exactly the same as the $\mathcal{H}\to 0$ limit studies of these systems \cite{omori_rheotaxis, BrethertonJFM1962}. Therefore, non-motile HT-S particles possess an
infinite family of steady-state heights. Notably, however, the field plays a crucial role in their orientational dynamics, where the channel separates into two distinct
regions, as illustrated by the swimmer schematics overlaid on the simulation
trajectories in Fig.~\ref{fig3}(a): far from the midplane, swimmers continuously
tumble, defining a \textbf{Tumbling region}, whereas near the midplane, their
orientations become stabilized, defining an \textbf{Oriented region}. The
distinct orientational dynamics in these regions are quantitatively captured
by the corresponding $\theta$--$t$ trajectories in Fig.~\ref{fig3}(b).

\smallskip

The trajectories in Fig.~\ref{fig3}(a) are shown for the initial orientation
$\theta_i=\pi$. The dynamics for general initial conditions $(y_i,\theta_i)$ are
summarized in the phase space shown in Fig.~\ref{fig3}(c), which reveals an
infinite number of stable and unstable Fixed Points (FPs) $(y^*,\theta^*)$ distributed
continuously (filled and open circles, respectively). However, these FPs
are bounded within a range of channel heights: swimmers initialized within this
range converge to a stable orientation specific to their initial height,
irrespective of $\theta_i$, whereas those initialized outside this range continuously
tumble. Thus, the channel is divided into horizontal layers separating the
\textbf{Oriented} and \textbf{Tumbling} regions, consistent with the vertically
separated \textbf{Oriented} and \textbf{Tumbling} states schematically depicted
in the $y$--$x$ trajectories of Fig.~\ref{fig3}(a).

\smallskip

\textbf{Dynamical Attractors and Their Stability.} We now quantitatively analyze the phase space dynamics described above. The $y$--dynamics of non-motile HT-S particles is given by Eq. \eqref{eq:ydotHTS_poiseuille}, with $v^{\rm{sp}}=0$, yielding $\dot{y}=0$ for all $y_i, \theta_i$. This explains the streamline motion of all initial conditions shown in Fig. \ref{fig3}(a). To obtain the steady state orientations, we look for the FPs of the system by requiring $\dot{\theta}=0$ along with $\dot{y}=0$. From Eq. \eqref{eq:thetadotHTS_poiseuille}, the condition $\dot{\theta}=0$ yields:
\begin{eqnarray}
y = \frac{\mathcal{H}\sin(\theta-\phi_0)}
{\frac{v_f}{R^2}\left[(1-G)+2G\sin^2\theta\right]}. \label{thetadot0nonmotileHTS}
\end{eqnarray}
The above is a $y(\theta)$ curve that provides the locus of all points in phase space where $\dot{\theta}=0$, known formally as the \textit{$\theta$--nullcline}. Since FPs require both $\dot{y}=0$ and $\dot{\theta}=0$, and $\dot{y}=0$ everywhere for non-motile HT-S particles, the entire $\theta$--nullcline therefore constitutes the locus of FPs for this particle class. Notably, the $\theta$--nullcline of Eq. \eqref{thetadot0nonmotileHTS} matches exactly with the numerically obtained locus of FPs in the phase space shown in Fig. \ref{fig3}(c).

\smallskip

To predict the stability of the FPs, we eliminate $y$ from Eqs. \eqref{eq:ydotHTS_poiseuille} and \eqref{eq:thetadotHTS_poiseuille}, reducing the coupled $y$--$\theta$ dynamics to a second-order equation for $\theta$ of the form $\ddot{\theta}=\ddot{\theta}(\theta,\dot{\theta})$. This maps the dynamics onto the $\theta$--$\dot{\theta}$ phase space. Linearizing about an FP $(y^*,\theta^*)\equiv(\theta^*,\dot{\theta}^*)$, with $\psi\equiv\theta-\theta^*$ and $\dot{\psi}\equiv\dot{\theta}-\dot{\theta}^*$, gives
\begin{eqnarray}
&&\ddot{\psi}+k^{\rm{HTS}}\psi=\zeta^{\rm{HTS}}_0\dot{\psi}.
\label{eq:stabilityHTS}
\end{eqnarray}
where $\zeta^{\rm HTS}_0= [2G \mathcal{H}\sin{(\theta^*-\phi_0)}\sin{2\theta^*}]/(1-G\cos{2\theta^*}) -\mathcal{H}\cos{(\theta^*-\phi_0)} $ and $k^{\rm{HTS}}=0$. Note that, when $\zeta^{\rm HTS}_0<0$, Eq. \eqref{eq:stabilityHTS} with $k^{\rm HTS}=0$ becomes an overdamped oscillator, with $\psi$--trajectories relaxing to $\psi\to0$, corresponding to the $\theta$--FP $(\theta^*)$. Therefore, those $\theta^*$ for which $\zeta^{\rm HTS}_0<0$ correspond to stable FPs, and coincide exactly with the filled circles in Fig. \ref{fig3}(c). On the other hand, those $\theta^*$ for which $\zeta^{\rm HTS}_0>0$ are unstable FPs, since Eq. \eqref{eq:stabilityHTS} then corresponds to a pumped oscillator, and these $\theta^*$ correspond exactly to the open circles in Fig. \ref{fig3}(c).

\smallskip

\textbf{Steady-State Orientations and Layering Mechanism.}
The stable FPs found above correspond to steady-state orientations $\theta^*$ which the HT-S particle ultimately adopts, irrespective of its initial orientation $\theta_i$, as exemplified by the orange trajectory in Fig. \ref{fig3}(b). However, the steady-state orientation depends on the swimmer height $y=y_i$, as governed by Eq. \eqref{thetadot0nonmotileHTS} and inferred from the FP locations in Fig. \ref{fig3}(c).

\smallskip

Crucially, our analysis predicts that the FP heights are bounded, as seen visually from the phase space in Fig. \ref{fig3}(c) and analytically from Eq. \eqref{thetadot0nonmotileHTS}. This equation is bounded by extremal heights $y^{\rm extreme}$, whose upper and lower bounds can be obtained by independently maximizing the numerator and minimizing the denominator of Eq. \eqref{thetadot0nonmotileHTS}, yielding the following sure bounds:
\begin{eqnarray}
    y^{\rm extreme}_{\pm}
    =\pm\frac{R^2}{v_f}\frac{\mathcal{H}}
    {(1-G)}.
\end{eqnarray}

\begin{figure*}
\centerline{\includegraphics[width=\linewidth]{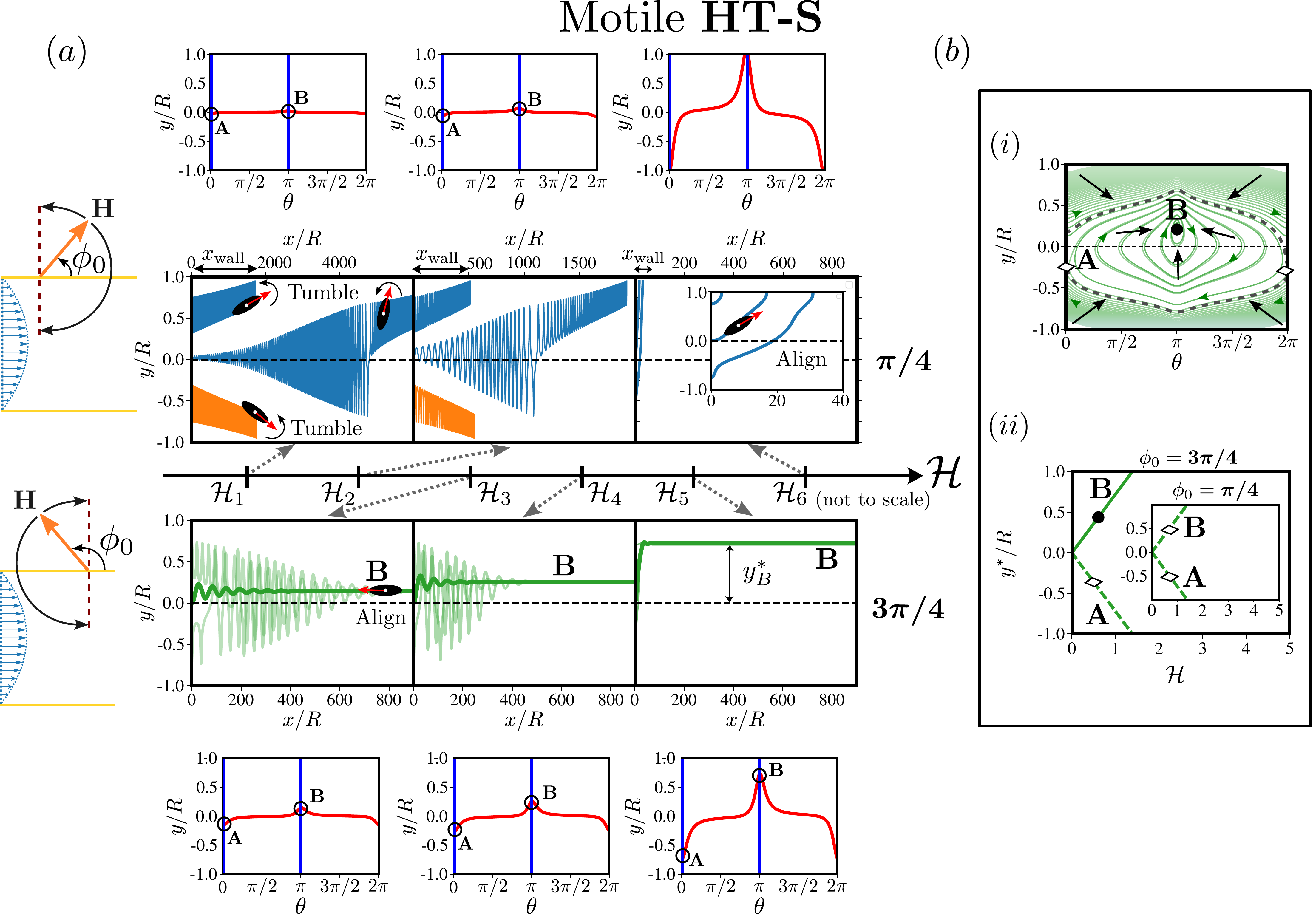}}
\caption{\textbf{Motile HT-S Particles} (a) Particle trajectories in channel flow for fields directed at $\phi_0=\pi/4$ (top row) and $\phi_0=3\pi/4$ (bottom row), with field-strengths: $\mathcal{H}_1=0.03\,\rm{s}^{-1}$, $\mathcal{H}_2=0.1\,\rm{s}^{-1}$, $\mathcal{H}_3=0.2\,\rm{s}^{-1}$, $\mathcal{H}_4=0.35\,\rm{s}^{-1}$, $\mathcal{H}_5=1\,\rm{s}^{-1}$ and $\mathcal{H}_6=1.5\,\rm{s}^{-1}$. The $y$--$\theta$ plots above and below the trajectory panels show the $\dot{y}=0$ (blue) and $\dot{\theta}=0$ (red) curves for field parameters $(\mathcal{H}, \phi_0)$ corresponding to the trajectories. Their intersections give the two fixed points (FPs) A and B. (b) -- (i) Phase space corresponding to $\mathcal{H}_4$ (with $\phi_0=3\pi/4$); the phase space for $\phi_0=\pi/4$ (not shown) is qualitatively similar to that of motile HT-A swimmers (see Fig. 2(d) in Ref. \cite{GomesAdhyapak_Field-Particle2026}). (ii) Heights $y^*$ of FPs A and B versus $\mathcal{H}$ for $\phi_0=\pi/4$; inset: $\phi_0=3\pi/4$. }

\label{fig4}
\end{figure*}

By mapping the above bounds on the FP heights to the boundaries of the Oriented
and Tumbling regions, we note that $y=y^{\rm extreme}_{+}$ approximates the
upper separatrix $\mathcal{S}_1$ curve, while $y=y^{\rm extreme}_{-}$
approximates the lower separatrix $\mathcal{S}_2$ curve. We have thus
quantitatively explained the horizontal layering of the channel, with the
\textbf{Oriented} region lying between the separatrices $\mathcal{S}_1$ and
$\mathcal{S}_2$, and the \textbf{Tumbling} region lying beyond them. Moreover,
Eq. \eqref{thetadot0nonmotileHTS} quantitatively captures how the steady-state
orientations vary with the field strength and orientation
$(\mathcal{H},\phi_0)$. Importantly, these variations produce only quantitative
changes: the horizontal layering into Oriented and Tumbling regions persists,
with only their relative extents changing. We therefore do not show the
corresponding variation with $\mathcal{H}$ and $\phi_0$ explicitly.

\smallskip

To our knowledge, such channel-layer-dependent particle orientations have not been reported previously for passive objects. Rather, in previous field--flow studies, the entire channel has been shown to host either tumbling or oriented states, with the state determined by the applied field strength (e.g., see \cite{Mason1974ParticleBehaviorShearElectricFields}).

The absence of such layering in previous studies can be attributed to their predominant consideration of flows with uniform velocity gradients. In contrast, the non-uniform shear in Poiseuille flow produces a height-dependent vorticity, which increases away from the channel midplane. Consequently, near the midplane, the vorticity-induced rotation is weaker and can be balanced by the field-induced rotation, allowing stable particle orientations. Further from the midplane, the increasing magnitude of the vorticity-induced rotation overwhelms the field-induced rotation, causing the particle to tumble about the vorticity direction. Thus, our analysis supports the broader conclusion that a particle experiencing an orienting field in a flow with non-uniform velocity gradients can exhibit distinct orientational dynamics in different regions of the channel.

\subsection{Motile HT-S Particles}

\textbf{Trajectory Classes and Dynamical Phase Space.} In the presence of motility, HT-S particles exhibit enhanced oscillatory dynamics, both with and without an applied field. The highly oscillatory trajectories in the $\mathcal{H}\to 0$ limit have been studied previously \cite{zottl_prl_NLD_channel, Benyahu_larva_in_flow}; here, we extend this analysis to field-oriented HT-S microswimmers. As seen from the $y$--$x$ trajectories in Fig. \ref{fig4}(a), the oscillations persist in the presence of a field, although the mean trajectory deviates from the $\mathcal{H}\to 0$ limit. In contrast to the non-motile case, the channel layering into regions with distinct orientational states now vanishes. Instead, two distinct trajectory classes emerge: \textbf{wall-focused} trajectories that approach the channel walls and \textbf{collimated} trajectories that converge to a unique, field-selected height. These trajectory classes depend strongly on the field direction $\phi_0$, as illustrated in Fig. \ref{fig4}(a): for $\phi_0$ in the first and fourth quadrants (top row), trajectories are exclusively wall-focused, whereas for $\phi_0$ in the second and third quadrants (bottom row), they are exclusively collimated.

While varying the field direction $\phi_0$ qualitatively changes the trajectory class, varying the field strength $\mathcal{H}$ provides quantitative control over the swimmer location and transport. For wall-focused trajectories with $\phi_0$ in the first quadrant, increasing $\mathcal{H}$ accelerates transport towards the wall, as indicated by the shorter $x$--distance travelled over the same $y$--distance in Fig. \ref{fig4}(a). Beyond a sufficiently strong field, the swimmer oscillations become suppressed [third panel in the top row, Fig. \ref{fig4}(a)], resulting in an extremely steep approach to the wall. For collimated trajectories with $\phi_0$ in the second quadrant, increasing $\mathcal{H}$ shifts the collimation location further away from the channel midplane.

\smallskip

Just as in the non-motile case, the $y$--$\theta$ phase space for $\phi_0$ in the second quadrant is shown in Fig. \ref{fig4}(b)-(i). The phase space contains a saddle FP A and a stable FP B. All trajectories approach FP B at long times, except those lying on the stable manifold of the saddle FP A, shown by the dashed gray curves. This stable manifold also forms the separatrix of the system, separating trajectories with unbounded orientations from those with bounded orientations [Fig. \ref{fig4}(b)-(i)]. The stable FP B corresponds to the collimated trajectory in the $y$--$x$ plane, with oscillations decaying to zero, while the separatrix divides trajectories that oscillate across the channel midplane from those that oscillate without crossing it. Notably, the phase space exhibits \textbf{broken horizontal layering} in contrast with the clear layers seen in the corresponding phase space of the non-motile HT-S class.

\smallskip

\textbf{Dynamical Attractors and Their Stability.} The FPs are points in phase space for which both $\dot y=0$ and $\dot{\theta}=0$. From Eq. \eqref{eq:ydotHTS_poiseuille}, the criterion $\dot y=0$ yields
$v^{\rm sp}\sin{\theta}=0$.
This implies that the locus of all points satisfying $\dot y=0$, i.e., the $y$--nullcline, is
\begin{eqnarray}
\theta=0,\pi ,\text{ for all }y.
\end{eqnarray}
This nullcline is independent of $y$ and therefore consists of two vertical lines in the $y$--$\theta$ phase space, shown as blue curves in Fig. \ref{fig4}(a). From Eq. \eqref{eq:thetadotHTS_poiseuille}, the criterion $\dot \theta=0$ yields the same $\theta$--nullcline as Eq. \eqref{thetadot0nonmotileHTS}, shown as a red curve in Fig. \ref{fig4}(a). The intersections of the $y$-- and $\theta$--nullclines are the FPs. We find that there are only two FPs, A and B, whose locations coincide with the FPs identified numerically as stable and saddle FPs, respectively.

To probe their stability, we once again obtain their angular $\psi$--dynamics from the phase-space Eqs. \eqref{eq:ydotHTS_poiseuille} and \eqref{eq:thetadotHTS_poiseuille}, following the procedure used for the non-motile case, but now with finite $v^{\rm sp}$. The corresponding $\psi$ equation has the form of Eq. \eqref{eq:stabilityHTS}, with $k^{\rm{HTS}}=-v_f v^{\rm sp}(1-G)/R^2$ and $\zeta^{\rm{HTS}}_0=-\mathcal{H}\cos{\phi_0}$ for FP A (where $\theta^*=0$), while $k^{\rm{HTS}}=-v_f v^{\rm sp}(1-G)/R^2$ and $\zeta^{\rm{HTS}}_0=+\mathcal{H}\cos{\phi_0}$ for FP B (where $\theta^*=\pi$).

\smallskip

We now use these expressions to determine the stability of the FPs for different field directions, starting with the \textit{second} and \textit{third} $\phi_0$--quadrants. In these quadrants, FP A has $k^{\rm{HTS}}<0$, corresponding to an oscillator with an inverted harmonic potential and hence unstable to generic deviations from the FP. Only the finely tuned initial conditions lying on its stable manifold, which exactly counterbalance the destabilizing effective potential, approach FP A, while all other trajectories are repelled from it. Thus, FP A is a \textit{saddle}, in agreement with the numerical phase space. In contrast, FP B has $k^{\rm{HTS}}>0$ and $\zeta^{\rm{HTS}}_0<0$, corresponding to a damped oscillator and hence a stable FP. Unlike the non-motile case, the finite $k^{\rm{HTS}}$, which depends on $v^{\rm sp}$, produces oscillations about FP B through the effective restoring term in the $\psi$ equation. The height of FP B follows directly from Eq. \eqref{thetadot0nonmotileHTS} by setting $\theta=\pi$:
\begin{eqnarray}
y^*_{B} = \frac{R^2}{v_f}\frac{\mathcal{H}\sin\phi_0}
{1-G}. \label{eq:ydotFPB}
\end{eqnarray}
Similarly, setting $\theta=0$ gives $y^*_{A}=-y^*_{B}$. These FP heights as functions of $\mathcal{H}$ are plotted in Fig. \ref{fig4}(b)-(ii) and show excellent agreement with the numerical results.

In the \textit{first} and \textit{fourth} $\phi_0$--quadrants, FP A remains a saddle since $k^{\rm{HTS}}<0$, while FP B becomes unstable because $k^{\rm{HTS}}>0$ and $\zeta^{\rm{HTS}}_0>0$, corresponding to a pumped oscillator. Although we do not show the corresponding phase space, our previous work \cite{GomesAdhyapak_Field-Particle2026} provides a phase-space analysis for $\phi_0=\pi/4$ that finds the same two FPs, one unstable and one saddle. While the phase space in that work was obtained for HT-A swimmers, at the field strength $\mathcal H=0.1\,\text{s}^{-1}$ considered there, the dynamics closely approach the corresponding HT-S dynamics, making it a relevant reference for comparison with our results. The FP heights for the first and fourth quadrants retain the same expressions as above; their variation with $\mathcal{H}$ for representative $\phi_0$ values in these quadrants is shown in the inset of Fig. \ref{fig4}(b)-(ii).

For field directions constrained to the flow axis, $\phi_0=0,\pi$, our results recover the findings of Ref.~\cite{salima_prl}. For an  upstream oriented field (corresponding to $\phi_0=\pi$), that work finds trajectories collimated only at the channel midplane, a location that cannot be modulated by the field strength $\mathcal{H}$, as follows directly by setting $\phi_0=\pi$ in Eq. \eqref{ydotFPB}. Their finding of wall-focused trajectories for downstream directed field ($\phi_0=0$) is likewise consistent with our results for the first and fourth $\phi_0$--quadrants.

\smallskip

\textbf{Origin of Field-Quadrant Sensitive Dynamics.}
The distinguishing feature of HT-S swimmers, their quadrant-selective dynamics, can be traced to the $\psi$--equation damping coefficient $\zeta_0^{\rm HTS}$ evaluated at FP B, whose form $\zeta^{\rm{HTS}}_0=-\mathcal{H}\cos{\phi_0}$ depends crucially on $\phi_0$. Changing $\phi_0$ from the second and third to the first and fourth quadrants therefore reverses the sign of $\zeta_0^{\rm HTS}$ and, consequently, the stability of FP B. Since a stable FP B acts as an attractor and produces collimated trajectories in the $y$--$x$ plane, whereas an unstable FP B does not, this change in stability ultimately gives rise to the qualitatively distinct dynamics observed across the different $\phi_0$--quadrants.

\smallskip

\textbf{Underlying Mechanism for Wall-focusing.}
For field directions $\phi_0$ in the first and fourth quadrants, there is no stable FP to stabilize the swimmer orientation and height, and hence the swimmer orientation tumbles. When the field has a vertical component, this tumbling is generally asymmetric: the swimmer rotates faster when oriented away from the field-side wall and more slowly when oriented towards it. It therefore spends more time facing the field-side wall, while self-propulsion simultaneously drives it towards that wall, producing the blue trajectories in Fig. \ref{fig4}(a). A small fraction of swimmers near the far-side wall exhibit the opposite asymmetry, rotating more slowly while oriented away from the field-side wall, and consequently reach the far-side wall instead [orange trajectories in Fig. \ref{fig4}(a)].

The steep wall-focusing observed at large $\mathcal{H}$ in the first and fourth quadrants arises because the field becomes strong enough to overcome the flow vorticity and approximately align the swimmer with the field direction [see schematic, Fig. \ref{fig4}(a), top row, third panel]. Once aligned, the swimmer is propelled towards the wall by its active velocity $v^{\rm{sp}}$, which then becomes the limiting factor for the rate of wall approach. Consequently, beyond a certain field strength, the wall-focusing rate is expected to saturate.

\subsection{Non-Motile HT-A Particles}
\label{subsec:nonmotileHTA}

\begin{figure*}
\centerline{\includegraphics[width=\linewidth]{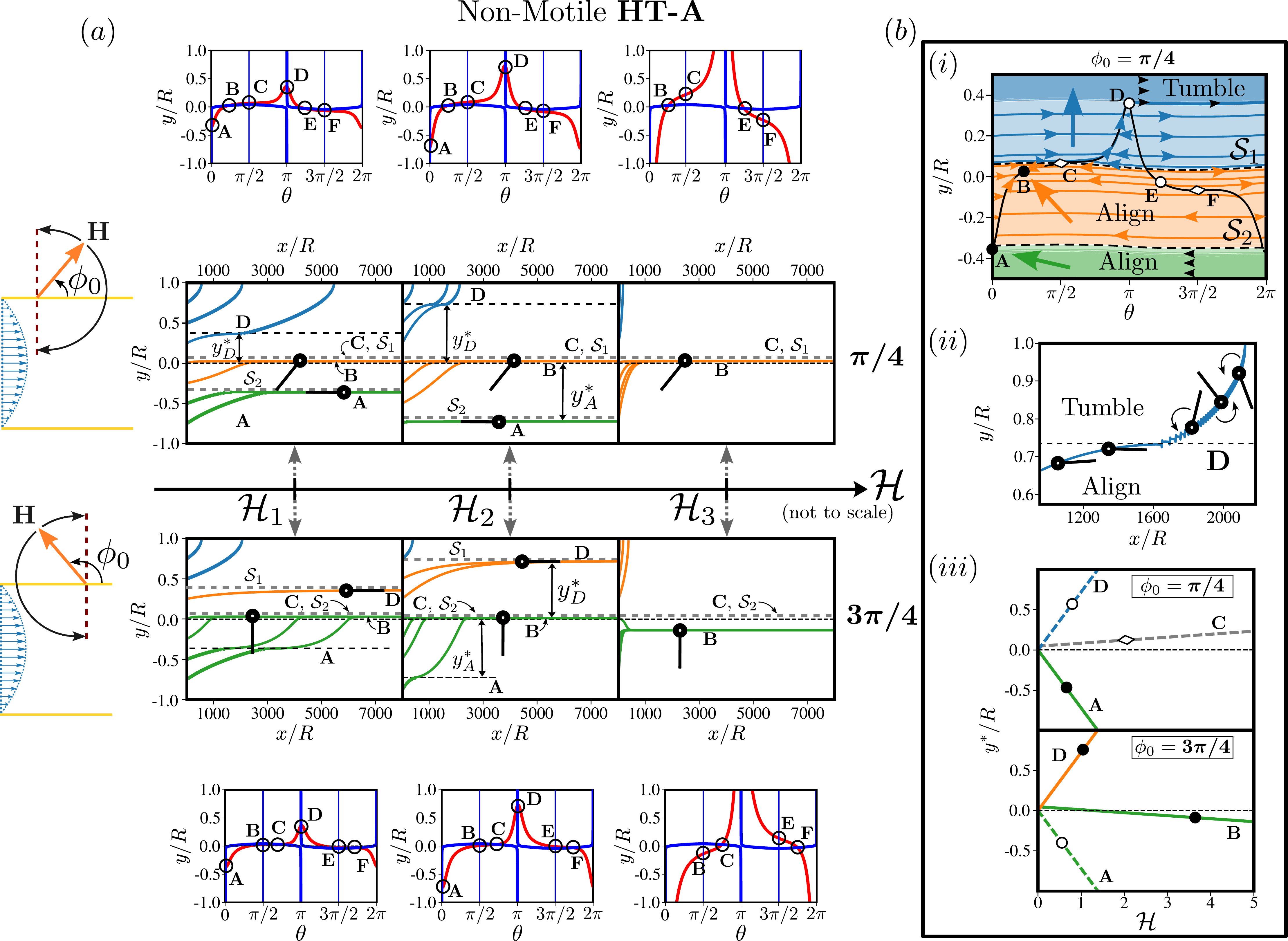}}
\caption{\textbf{Non-motile HT-A Particles:} (a) $y$--$x$ and $y$--$\theta$ plots for fields directed at $\phi_0=\pi/4$ (top) and $\phi_0=3\pi/4$ (bottom), as in Fig. \ref{fig4}(a), with $\mathcal{H}_1=0.5\,\rm{s}^{-1}$, $\mathcal{H}_2=1.0\,\rm{s}^{-1}$ and $\mathcal{H}_3=5.0\,\rm{s}^{-1}$. The six fixed points (FPs) obtained from the intersections of the curves in the $y$--$\theta$ plots are labelled A--F. (b) -- (i) Phase space for $\phi_0=\pi/4$ and $\mathcal{H}_1$, containing the FPs A--F: stable FPs (filled circles), unstable FPs (open circles), and saddle FPs (open diamonds), together with separatrices $\mathcal{S}_1$ and $\mathcal{S}_2$. (ii) Plot of $y$-$x$ near $y_D^*$ for parameters ($\mathcal{H}_2,\pi/4$), showing the transition from approximately aligned to continuously tumbling particle orientations. (iii) Heights of the relevant FPs versus $\mathcal{H}$ obtained analytically for $\phi_0=\pi/4$ (top) and $\phi_0=3\pi/4$ (bottom).}

\label{fig5} 
\end{figure*}

With HT-A, even non-motile particles exhibit nontrivial cross-stream dynamics. We first study the $\mathcal{H}\to 0$ limit of this particle class, for varying initial particle heights $y_i$, shown in Fig. \ref{fig4}(d). Remarkably, the cell-body centre $\rb_s$ does not follow a streamline, but instead undergoes oscillations about a mean trajectory that follows the streamline on average, as seen from the blue trajectories in the figure. These oscillatory trajectories closely resemble experimentally reported trajectories of non-motile \textit{T. brucei} in a channel \cite{Stark_blood_flow}. Our results therefore suggest that the oscillations observed experimentally can arise from the HT-A effects of \textit{T. brucei}, which is indeed shape asymmetric.

\smallskip

Figure~\ref{fig3}(d) also shows a small region near the channel midplane, with trajectories shown in orange, that are focused to nearby locations. However, this region is negligibly small for the parameters considered here and is therefore not investigated further. We now turn to the dynamics in the presence of an external field.

\smallskip

\textbf{Trajectory Classes and Dynamical Phase Space.} In the presence of a field, the dynamics of non-motile HT-A particles becomes much richer, as captured by the $y$--$x$ trajectories in Fig. \ref{fig5}(a). The particle trajectories are no longer confined to streamlines even on average, but instead cross them, leading to trajectories with \textbf{multiple collimation} locations or trajectories that are \textbf{wall-focused}, all despite being non-motile. Unlike the HT-S swimmer case, however, the dynamics is no longer strongly quadrant-selective. Instead, varying $\mathcal{H}$ markedly alters the dynamics, changing both the number and distance from the midplane of the collimation locations, while also modifying the rate of approach to the walls for wall-focused trajectories, as shown in Fig. \ref{fig5}(a). Notably, for a given field strength and orientation $(\mathcal{H},\phi_0)$, the initial particle height crucially determines the resulting trajectory state. This is evident in Fig. \ref{fig5}(a), where trajectories released from different $y_i$ with the same $\theta_i=\pi$ reach different states, demonstrating the strong dependence of the trajectory state on $y_i$.

\smallskip

While the above trajectories have a fixed $\theta_i$, we construct the $y$--$\theta$ phase space for this particle class as well, shown in Fig. \ref{fig5}(b)-(i), thereby revealing the dynamics for all initial conditions $(y_i,\theta_i)$. The phase space contains multiple dynamical attractors: \textit{six FPs}, labelled A--F, and \textit{two separatrices}, $\mathcal{S}_1$ and $\mathcal{S}_2$. These separatrices divide the phase space into three \textbf{horizontal layers}, coloured blue, orange, and green. In the blue layer, trajectories continuously tumble and ultimately reach the channel boundaries, corresponding to the wall-focused trajectories in the $y$--$x$ plane, also shown in blue. In the orange layer, trajectories approach the stable FP B, corresponding to the orange $y$--$x$ trajectories collimated at a particular height within the channel. Similarly, trajectories in the green layer approach the stable FP A, corresponding to the green $y$--$x$ trajectories collimated at another, lower height. Since the separatrices $\mathcal{S}_1$ and $\mathcal{S}_2$ are approximately horizontal, their heights vary negligibly with $\theta$. Consequently, the initial height $y_i$ predominantly determines the layer to which a particle belongs and hence its eventual dynamical state. This explains the strong dependence of the trajectory state on the initial height $y_i$ observed in the $y$--$x$ trajectories of Fig. \ref{fig5}(a).

\smallskip

While the stable FPs A and B have been shown to crucially control the particle dynamics, the saddle FP C and unstable FP D also play important roles. The stable manifold of the saddle FP C forms the separatrix $\mathcal{S}_1$, so its location determines the boundary between the blue and orange layers, corresponding to the boundary between wall-focused and collimated trajectories in the $y$--$x$ plane. The unstable FP D, in contrast, marks the boundary between unbounded particle orientations (tumbling) and bounded orientations. This is evident in the phase space in Fig. \ref{fig5}(b)-(i) and manifests as an abrupt onset of oscillations in the $y$--$x$ trajectories beyond the height marked D in Fig. \ref{fig5}(b)-(ii). This height coincides exactly with the height of the unstable FP D.

\smallskip

\textbf{Dynamical Attractors and Their Stability.} Using similar techniques as for the other particle classes, we now quantitatively analyse the numerically observed phase-space dynamics. The FPs are points in phase space for which both $\dot y=0$ and $\dot{\theta}=0$. For non-motile HT-A particles, Eq. \eqref{eq:ydotHTA_poiseuille} with $v^{\rm sp}=0$ gives the condition $\dot y=0$ as
\begin{eqnarray}
\dot{y}&=&
r_{\rm{sh}}\dot{\theta}\cos\theta
+\dot{y}^{\rm{HTS}}_{\rm{flow}}
+\dot{y}^{\rm{HTA}}_{\rm{flow}}=0.
\label{eq:ydotHTA=0}
\end{eqnarray}
Substituting $\dot{\theta}=0$ into Eq. \eqref{eq:ydotHTA=0}, together with the expressions for $\dot{y}^{\rm{HTS}}_{\rm{flow}}$ and $\dot{y}^{\rm{HTA}}_{\rm{flow}}$ from Eqs. \eqref{expr:ydotHTS_flow} and \eqref{expr:ydotHTA_flow}, respectively, gives
\begin{eqnarray}
y&=&\tilde{y}_1(\theta)  =\frac{
-\frac{a^2}{3}
+\left(a^2+a\ell_f+\frac{\ell_f^2}{2}\right)\sin^2\theta}
{2(a+\ell_f)\sin\theta}.
\label{eq:ytildenullcline}
\end{eqnarray}
The expression $\tilde{y}_1(\theta)$ therefore gives the locus of points satisfying $\dot y=0$ subject to $\dot{\theta}=0$. However, it is undefined at $\theta=0,\pi/2,\pi,3\pi/2$, where its denominator vanishes. Direct substitution of these orientations into Eq. \eqref{eq:ydotHTA_poiseuille} shows that, for non-motile particles ($v^{\rm sp}=0$), all four orientations give $\dot y=0$ for any $y$. Thus, the complete locus satisfying $\dot y=0$ subject to $\dot{\theta}=0$ is
\begin{eqnarray}
    y&=&y_1(\theta)\nonumber\\
     &=&\tilde{y}_1(\theta)\,\text{ and all }y\text{ at }
    \theta=0,\frac{\pi}{2},\pi,\frac{3\pi}{2}.
    \label{eq:ynullcline}
\end{eqnarray}
This locus is plotted in blue in the $y$--$\theta$ plots in Fig. \ref{fig5}(a). We next determine the locus of points satisfying $\dot{\theta}=0$. Setting $\dot{\theta}=0$ in Eq. \eqref{eq:thetadotHTA_poiseuille} gives
\begin{eqnarray}
y&=&y_2(\theta)=
\frac{\mathcal{H}\sin(\theta-\phi_0)
-\frac{v_f}{R^2}\left[G_1\sin\theta+G_2\sin^3\theta\right]}
{\frac{v_f}{R^2}\left[1-G+2G\sin^2\theta\right]}.
\label{eq:thetanullcline}
\end{eqnarray}
This curve is plotted in red in the $y$--$\theta$ plots in Fig. \ref{fig5}(a). Notably, the line connecting all the FPs in the phase space of Fig. \ref{fig5}(b)--(i) exactly matches $y_2(\theta)$. The intersections of the loci $y_1(\theta)$ and $y_2(\theta)$ therefore give the points satisfying both $\dot y=0$ and $\dot{\theta}=0$, and hence the FPs of the system. We find exactly six such intersections in the $y$--$\theta$ phase space of Fig. \ref{fig5}(a), which we now show coincide exactly with the FPs identified numerically in Fig. \ref{fig5}(b)-(i).

\smallskip

To study the stability of all the FPs identified above, we consider the HT-A phase-space dynamics governed by Eqs. \eqref{eq:ydotHTA_poiseuille} and \eqref{eq:thetadotHTA_poiseuille}. Following the same procedure used for the HT-S case, we eliminate $y$ to obtain the corresponding $\psi$--equation:
\begin{eqnarray}
&&\ddot{\psi}+k^{\rm{HTA}}\psi=\zeta^{\rm{HTA}}_0\dot{\psi}.
\label{eq:stabilityHTA}
\end{eqnarray}
Here, $k^{\rm{HTA}}$ and $\zeta^{\rm{HTA}}_0$ are obtained in terms of the known parameters and the FP orientation $\theta^*$. As in the HT-S case, the classification of the FPs follows from the signs of these two quantities: $k^{\rm{HTA}}>0$ with $\zeta^{\rm{HTA}}_0<0$ implies a stable FP, $k^{\rm{HTA}}>0$ with $\zeta^{\rm{HTA}}_0>0$ implies an unstable FP, while $k^{\rm{HTA}}<0$ implies a saddle FP. Applying these criteria, we find that for $\phi_0=\pi/4$, FPs A and B are stable, D and E are unstable, and C and F are saddles. For $\phi_0=3\pi/4$, FPs B and D are stable, A and E are unstable, and C and F are saddles. The FPs are labelled A--F in increasing order of their $\theta^*$ values [see Fig. \ref{fig5}(a)]. The nature of the FPs determined from this analysis matches exactly with the simulation results, as seen for the $\phi_0=\pi/4$ case in the phase space plotted in Fig. \ref{fig5}(b)-(i).

\smallskip

As discussed above, for $\phi_0=\pi/4$, the stable FPs A and B, the unstable FP D, and the saddle FP C have important impacts on the particle trajectories. For analogous reasons, for $\phi_0=3\pi/4$, the stable FPs B and D and the unstable FP A likewise impact the particle trajectories. We now show that the variation of the heights of some of these FPs with increasing $\mathcal{H}$ can be determined analytically in a straightforward manner.

To determine this dependence, we note that the orientations $\theta^*=0,\pi/2,\pi$ and $3\pi/2$ are FPs of the system irrespective of $\mathcal{H}$. For $\phi_0=\pi/4$, these orientations correspond to FPs A, C, D and F, while for $\phi_0=3\pi/4$, they correspond to A, B, D and E [see $y$-$\theta$ plots, Fig. \ref{fig5}(a)]. Their heights can be obtained directly from Eq. \eqref{eq:thetanullcline} of the $\theta$--nullcline (where $\dot \theta=0$), by substituting their orientations, yielding:
\begin{eqnarray}
y^*&=&
\frac{\mathcal{H}\sin(\theta^*-\phi_0)
-\frac{v_f}{R^2}\left[G_1\sin\theta^*
+G_2\sin^3\theta^*\right]}
{\frac{v_f}{R^2}\left[1-G+2G\sin^2\theta^*\right]}.
\label{eq:ystar_vs_H}
\end{eqnarray}
Equation \eqref{eq:ystar_vs_H} thus gives the field-strength dependence of their heights, plotted for both $\phi_0=\pi/4$ and $\phi_0=3\pi/4$ in Fig. \ref{fig5}(b)-(iii) for the FPs most relevant to the particle dynamics, in excellent agreement with our numerical observations.

\smallskip

\textbf{Underlying mechanisms behind Collimation and Wall-focusing:} 

\smallskip

\textbf{(a) Multiple Collimation Heights}. A remarkable result of our study is that a single applied field can produce multiple focusing heights. We show that this arises as a consequence of three distinct mechanisms of cross-stream arrest, available to the system. The first occurs when the field and flow orient the particle parallel or antiparallel to the channel, such that $\sin\theta^*=0$. At these orientations, both flow-induced cross-stream contributions vanish, $\dot{y}^{\rm HTS}_{\rm flow}=\dot{y}^{\rm HTA}_{\rm flow}=0$ [see Eqs. \eqref{expr:ydotHTS_flow} and \eqref{expr:ydotHTA_flow}]. This produces the stable FP A for $\phi_0=\pi/4$ and FP D for $\phi_0=3\pi/4$.

The second mechanism occurs when the particle settles at a non-horizontal orientation, for which the HT-S and HT-A contributions to cross-stream motion are individually nonzero but exactly balance, yielding $\dot{y}^{\rm HTS}_{\rm flow}+\dot{y}^{\rm HTA}_{\rm flow}=0$. This balance occurs at a specific height because $\dot{y}^{\rm HTA}_{\rm flow}$ couples to the spatially varying flow gradient and therefore varies across the channel, allowing it to cancel $\dot{y}^{\rm HTS}_{\rm flow}$ at a particular height. This mechanism produces FP B for $\phi_0=\pi/4$.

The third mechanism occurs for FP B when $\phi_0=3\pi/4$, where the particle settles perpendicular to the flow. Here, $\cos{\theta^*}=0$, causing both $\dot{y}^{\rm HTS}_{\rm flow}$ and $\dot{y}^{\rm HTA}_{\rm flow}$ to once again vanish identically [see Eqs. \eqref{expr:ydotHTS_flow} and \eqref{expr:ydotHTA_flow}], arresting cross-stream motion. Thus, multiple focusing heights can emerge from a single applied field because HT-A dynamics admits multiple, physically distinct routes to cross-stream arrest.

\smallskip

\textbf{(b) Wall-focusing.} In Fig. \ref{fig5}(b)-(ii) for $\phi_0=\pi/4$, as the particle approaches FP D at height $y^*_{D}$, the competing flow- and field-induced rotations tend to align the particle upstream with the flow, thereby suppressing its cross-stream motion. Beyond $y^*_{D}$, however, the increasing shear towards the channel walls causes the flow-induced rotation to overwhelm the field-induced rotation, so the particle can no longer maintain this upstream orientation and instead tumbles continuously, as seen above height $y^*_{D}$ in Fig. \ref{fig5}(b)-(ii). For HT-A particles, this loss of flow alignment restores cross-stream motion through the coupling of asymmetric particle rotation rates to the flow-gradient-induced translation $\dot{y}^{\rm HTA}_{\rm flow}$ in Eq. \eqref{eq:ydotHTA_poiseuille}. The particle consequently crosses streamlines and is ultimately driven towards the channel wall.

\smallskip
We note that a key player in both the collimation and wall-focusing dynamics is $\dot{y}^{\rm HTA}_{\rm flow}$, which originates from the HT-A polar force that generates the $\Gammab_1$ term in Eq. \eqref{eq:vh_flow}. The HT-A torques that generate the $\kappa_{2,3}$ terms in Eq. \eqref{eq:omegah_flow}, on the other hand, give rise to the $G_{1,2}$ terms in the $\theta$--nullcline, Eq. \eqref{eq:thetanullcline}, which determine several of the FP locations.

\subsection{Motile HT-A Particles}
\label{subsec:motileHTA}

\begin{figure*}
\centerline{\includegraphics[width=\linewidth]{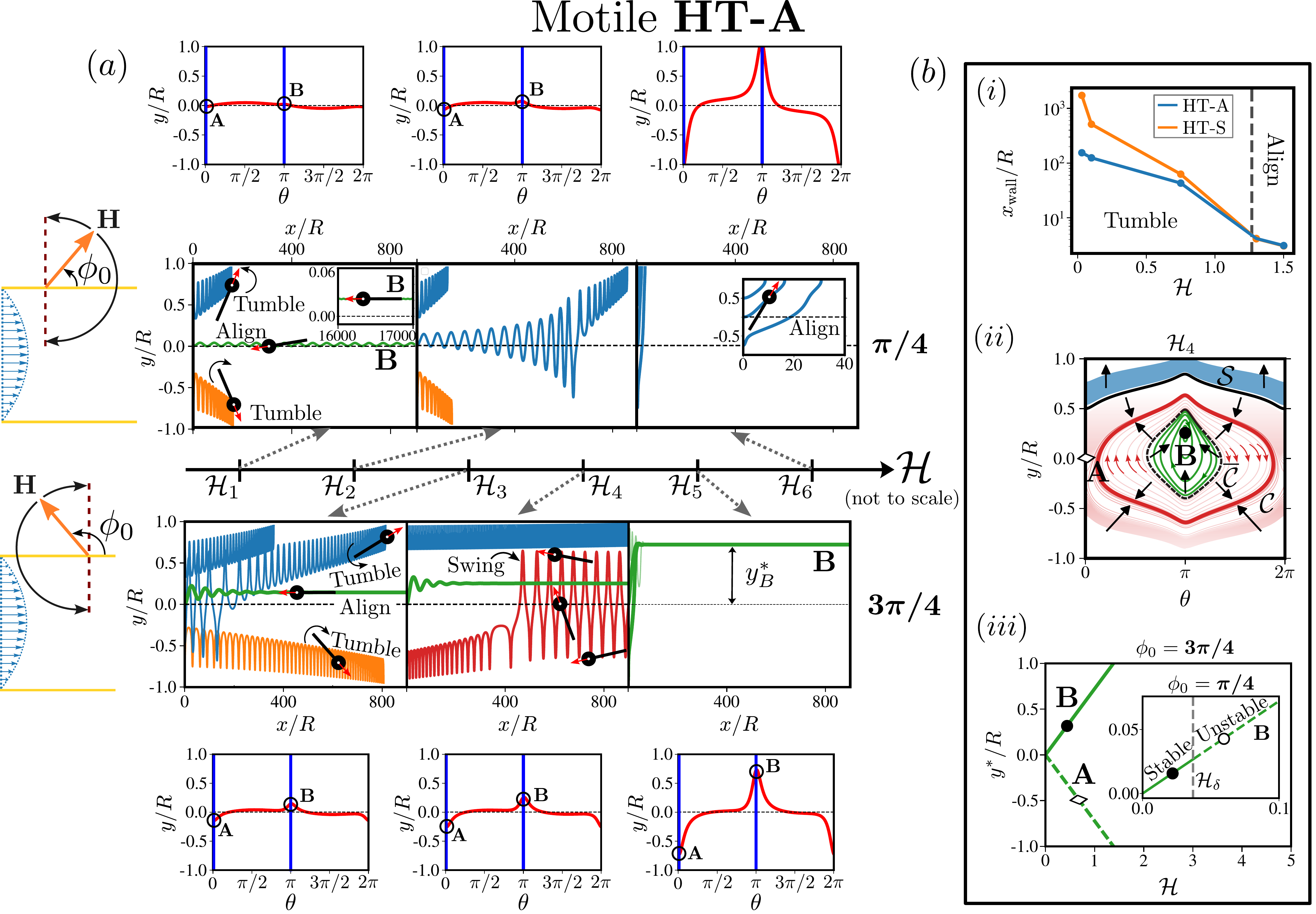}}
\caption{\textbf{Motile HT-A Particles} (a) Swimmer trajectories and corresponding $y$--$\theta$ plots for varying $\mathcal{H}$ and $\phi_0$, using the same parameter values as the corresponding panels in Fig. \ref{fig4}(a). The intersections of the $y$--$\theta$ curves give the FPs A and B. (b) -- (i) Downstream distance $x_{\rm wall}$ traveled by HT-S and HT-A swimmers to reach $y_i=0.95R$, starting from $y_i=0.75R$ and $\theta_i=\pi$, as a function of $\mathcal{H}$. (ii) Phase space for $\mathcal{H}_4$ ($\phi_0=3\pi/4$), containing stable and unstable FPs A,  B, stable and unstable LCs $\mathcal{C}$, $\overline{\mathcal{C}}$, and separatrices $\mathcal{S}$, $\overline{\mathcal{S}}$. FP symbols denote their respective stability types, as in Fig. \ref{fig5}(b)-(i). (iii) FP heights $y^*$ versus $\mathcal{H}$ from analytics for $\phi_0=3\pi/4$. Inset: Corresponding $y_B^*$ versus $\mathcal{H}$ for $\phi_0=\pi/4$, showing the stability transition at $\mathcal{H}_{\delta}$.}

\label{fig6}
\end{figure*}

Microswimmers such as \textit{E. coli} and \textit{B. subtilis}, as well as their synthetic analogs, which are widely studied in active matter research, fall under the class of motile HT-A microswimmers owing to their head--tail asymmetry and motility. The $\mathcal{H}\to0$ limit of this class was studied in Ref. \cite{GomesAdhyapakPRL2025}, where cross-stream migration was found to be governed by an unstable limit cycle, giving rise to oscillatory trajectories with either growing amplitude, leading to \textbf{wall-focused} states, or decaying amplitude, leading to \textbf{collimation} at the channel midplane.

\smallskip

\textbf{Trajectory Classes and Dynamical Phase Space.} In the presence of a field, the trajectories of HT-A swimmers are shown in Fig. \ref{fig6}(a). The trajectories continue to exhibit \textbf{wall-focused} (blue/orange trajectories) and \textbf{collimated} (green trajectories) states, with the steepness of wall-focusing and the plane of collimation controlled by the field strength $\mathcal{H}$, as in the HT-S case, though with differing rates. In addition, a \textbf{stable oscillatory} (red trajectory) state emerges at intermediate $\mathcal{H}$ that is unique to this particle class. In this state, swimmer trajectories are collimated into an extended beam formed by oscillations with a unique, field- and flow-selected amplitude and mean location.

At weak to intermediate $\mathcal{H}$, motile HT-A trajectories differ qualitatively and quantitatively from their HT-S counterparts, as seen in Fig. \ref{fig6}(a). The quantitative difference is captured in the $x_{\rm wall}$--$\mathcal{H}$ plot in Fig. \ref{fig6}(b)-(i), where the wall-focusing lengths of the two particle classes differ drastically at weak fields, as evident from the logarithmic scale of the plot. At large $\mathcal{H}$, however, the HT-A trajectories collapse onto those of their HT-S counterparts, as seen in the rightmost panels of Figs. \ref{fig4}(a) and \ref{fig6}(a). In this limit, $x_{\rm wall}$ for the two particle classes also converges, as seen in Fig. \ref{fig6}(b)-(i).

\smallskip

The $y$--$\theta$ phase space of this system provides the dynamics for all initial conditions and houses the dynamical attractors underlying the observed trajectory classes. The phase space for $\phi_0=3\pi/4$ is plotted in Fig. \ref{fig6}(b)-(ii). It contains two FPs, a saddle FP A and a stable FP B, as well as a stable and unstable limit-cycle (LC) pair, $\mathcal{C}$ and $\overline{\mathcal{C}}$, respectively, and a separatrix $\mathcal{S}$. The phase space is partitioned into three main regions. The green region enclosed by $\overline{\mathcal{C}}$ is the basin of attraction of FP B; all trajectories within this region approach FP B. The red region outside $\overline{\mathcal{C}}$ and within $\mathcal{S}$ is the basin of attraction of $\mathcal{C}$; all trajectories within this region approach $\mathcal{C}$. The blue region outside $\mathcal{S}$ contains trajectories that reach the $y$--boundary, i.e. the channel wall at $y=R$.

These three regions map directly onto the three trajectory classes in the $y$--$x$ plane shown in the bottom row of Fig. \ref{fig6}(a). The entire green basin of FP B corresponds to the green $y$--$x$ trajectories that become collimated, the entire red basin of $\mathcal{C}$ corresponds to the red trajectories that settle into oscillations of fixed amplitude, and the blue region outside $\mathcal{S}$ corresponds to the blue trajectories that approach the channel wall. For $\phi_0=\pi/4$, the corresponding phase-space structure, analysed in Ref. \cite{GomesAdhyapak_Field-Particle2026}, changes with $\mathcal{H}$. At intermediate to large $\mathcal{H}$, FP B is unstable and trajectories instead approach the channel walls, corresponding to the blue and orange trajectories in the top row of Fig. \ref{fig6}(a), whereas at weak $\mathcal{H}$, FP B becomes stable and is surrounded by an unstable LC that bounds its basin of attraction, giving rise to the green collimated trajectories in the top row of Fig. \ref{fig6}(a).

\textbf{Broken Layering.} Notably, unlike the prominent horizontal layering of the non-motile HT-A class, motility dissolves this horizontal layering, as it did for the motile HT-S class. This broken horizontal layering is clearly manifested in the phase space of Fig. \ref{fig6}(b)-(ii), in contrast with the clear layers seen in the corresponding non-motile HT-A phase space [Fig. \ref{fig5}(b)-(i)]. Consequently, the trajectory behaviour is no longer determined solely by the initial height $y_i$, but also depends on the initial orientation $\theta_i$.

\smallskip

\begin{figure}
\centerline{\includegraphics[width=\columnwidth]{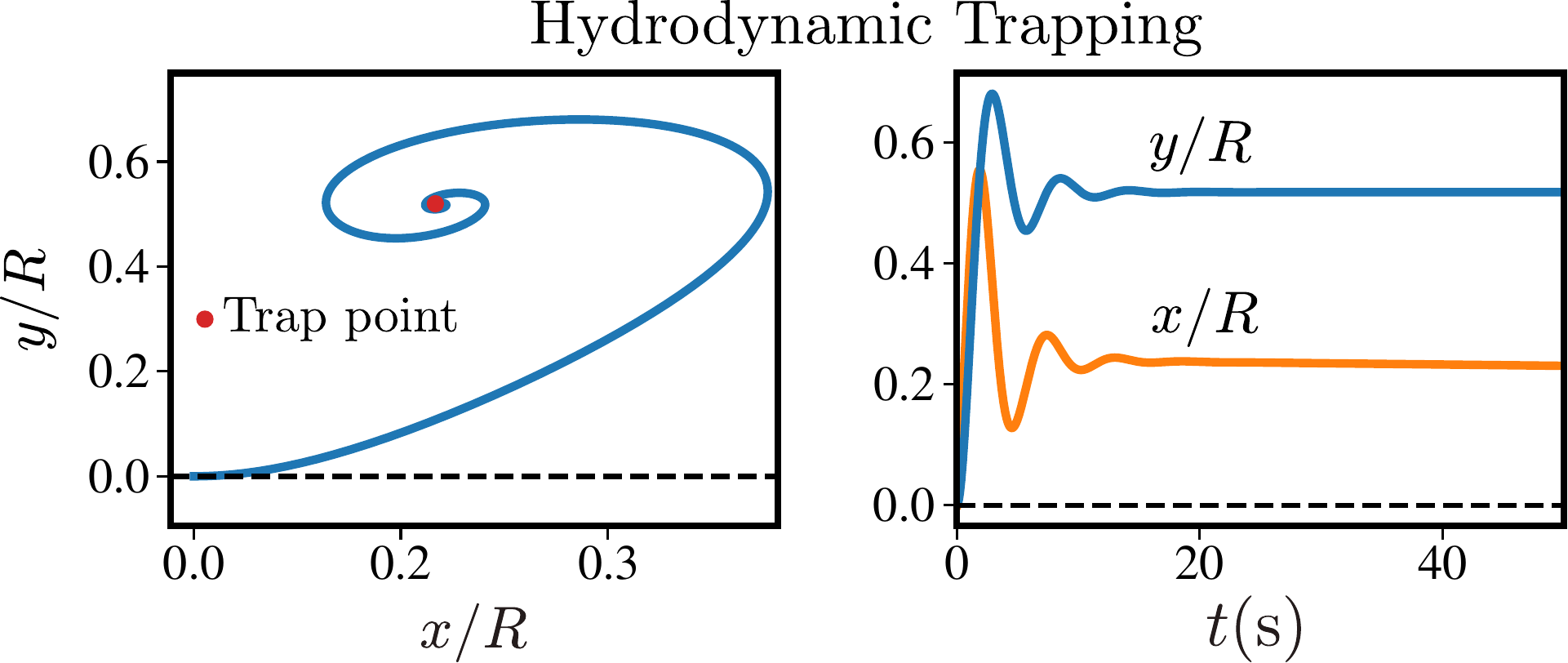}}
\caption{(a) $y$--$x$ trajectory terminating at a fixed spatial point (orange dot) for flow $v^*_f=68.30\,\mu\text{m}/\text{s}$, $\mathcal{H}=1\,\rm{s}^{-1}$, and $\phi_0=3\pi/4$. (b) Corresponding $x/R$ and $y/R$ trajectories versus $t$, showing positional saturation after $\approx15\,\text{s}$ and hence arrested motion.}

\label{fig7}
\end{figure}

\textbf{Dynamical Attractors and Their Stability.} We now quantitatively analyze the phase-space dynamics. As noted before, the FPs are the $y$--$\theta$ points that satisfy both $\dot{y}=0$ and $\dot{\theta}=0$. Setting $\dot{y}$ in Eq. \eqref{eq:ydotHTA_poiseuille} to zero gives the criterion for $\dot{y}=0$ for HT-A swimmers:
\begin{eqnarray}
    \dot{y}&=&
    v^{\rm sp} \sin{\theta}+ r _{\rm{sh}}\dot{\theta}\cos\theta
    +\dot{y}^{\rm{HTS}}_{\rm{flow}}
    +\dot{y}^{\rm{HTA}}_{\rm{flow}}=0.
    \label{eq:ydotHTAmotile=0}
\end{eqnarray}
For the motility considered here, $v^{\rm sp}=50\,\mu\text{m}/\text{s}$, the self-propulsion term is sufficiently large that it cannot be counterbalanced by the remaining contributions to $\dot{y}$ within the channel. Thus, $\dot{y}=0$ requires all the terms in $\dot{y}$ to vanish individually. This requires $\theta=0$ or $\pi$. At these orientations, $v^{\rm sp}\sin{\theta}=0$, while $\dot{y}^{\rm HTS}_{\rm flow}$ and $\dot{y}^{\rm HTA}_{\rm flow}$ also vanish [see Eqs. \eqref{expr:ydotHTS_flow} and \eqref{expr:ydotHTA_flow}]. Moreover, the $r_{\rm{sh}}$ term vanishes when $\dot{\theta}=0$. The latter condition is given by the $\theta$-nullcline, Eq. \eqref{eq:thetanullcline}, which is identical to that for non-motile HT-A swimmers. Hence, the two $\dot{y}=0$ branches subject to $\dot{\theta}=0$, $\theta=0$ and $\pi$, together with the $\theta$-nullcline form the nullclines of the system, shown in the $y$--$\theta$ plots in Fig. \ref{fig7}(a) as blue and red curves, respectively. Their intersections satisfy $\dot{y}=\dot{\theta}=0$, giving the two FPs, A and B, which correspond to the two numerically obtained FPs in the phase space of Fig. \ref{fig6}{b}-(ii). The heights of these two FPs as functions of $\mathcal{H}$ are found from Eq. \eqref{eq:thetanullcline} upon substituting $\theta=0,\pi$, yielding the heights $y^*_{A}$ and $y^*_{B}$ of FPs A and B, respectively:
\begin{eqnarray}
    y^*_{A} &=& \frac{-\mathcal{H}\sin\phi_0}
    {\frac{v_f}{R^2}(1-G)}, \qquad
    y^*_{B} = \frac{\mathcal{H}\sin\phi_0}
    {\frac{v_f}{R^2}(1-G)}.
\end{eqnarray}
These FP heights for $\phi_0=3\pi/4$ are plotted in Fig. \ref{fig6}(b)-(iii).

\smallskip

The stability of the FPs follows, as for the earlier particle classes, from the $\psi$--equation of the same form as in Eq. \eqref{eq:stabilityHTA}, but for the case with finite $v^{\rm{sp}}$. For the motile HT-A case, the coefficients then take the form:
\begin{eqnarray}
k^{\rm{HTA}}&=&-\cos{\theta^*}\left( 1-G \right)\frac{v_f}{R^2}
\left[v^{\rm{sp}}-\left(\frac{v_f}{R^2}\right)\frac{a^2\mathcal{A}_y}{3}\right],\\
\zeta^{\rm{HTA}}_0&=&\cos{\theta^*}\left\{\mathcal{H}\cos{\phi_0}
-\frac{v_f}{R^2}\left[G_1+\left(1-G\right)r_{\rm{sh}}\right]\right\}.
\end{eqnarray}
FP A ($\theta^*=0$) is therefore always a saddle, since $k^{\rm{HTA}}<0$. FP B ($\theta^*=\pi$), in contrast, is stable for $\phi_0$ in the second and third quadrants, where $\zeta^{\rm{HTA}}_0<0$. In the first and fourth quadrants, $\zeta^{\rm{HTA}}_0$ can change sign, allowing the stability of FP B to switch. As an illustration, for $\phi_0=\pi/4$ (first quadrant), the field strength $\mathcal{H}_{\delta}$ at which this transition occurs is obtained by setting $\zeta^{\rm{HTA}}_0=0$, yielding
\begin{eqnarray}
\mathcal{H}_{\rm{\delta}} = \frac{1}{\sqrt{2}}\frac{v_f}{R^2}
\left[G_1+\left(1-G\right)r_{\rm{sh}}\right].
\label{expr:Hdelta}
\end{eqnarray}
For our parameters, $\mathcal{H}_{\delta}\approx0.037,\rm{s}^{-1}$, as demarcated in the inset of Fig. \ref{fig6}(b)-(iii). This stability transition explains the green stable trajectory for $\mathcal{H}_1$ in Fig. \ref{fig6}(a), for which $\phi_0=\pi/4$ and $\mathcal{H}<\mathcal{H}_{\delta}$, and its disappearance for $\mathcal{H}>\mathcal{H}_{\delta}$ in $\mathcal{H}_2$ and $\mathcal{H}_6$.

\smallskip

\textbf{Role of Large Activity and HT-A Torques.} For motile HT-A swimmers, self-propulsion dominates the $\dot y$--dynamics, masking the flow-induced contributions. As a result, the additional FPs arising from the competition between the various flow-induced contributions, $\dot{y}^{\rm HTS}_{\rm flow}$ and $\dot{y}^{\rm HTA}_{\rm flow}$, disappear. Thus, HT-A microswimmers do not introduce additional steady-state heights beyond those of motile HT-S swimmers, in contrast to the non-motile HT-A case. In the $\dot\theta$ dynamics, however, the field-induced and flow-curvature-induced HT-A torques are comparable, giving rise to the rich HT-A dynamics when coupled to swimmer activity, including the LC attractors that greatly modify both the phase space and the trajectory dynamics. The mechanisms behind the formation of these LCs have been presented in our previous work \cite{GomesAdhyapak_Field-Particle2026}, while here we have exhaustively searched for and identified all possible FPs, determined their stability, and mapped their effects onto the corresponding $y$--$x$ trajectories. Furthermore, we have provided a quantitative analysis of swimmer dynamics across varying field strengths $\mathcal{H}$ and directions $\phi_0$.

\subsection{Applications: Particle Trapping, Sorting and Phase Control}
\label{subsec:Applications}

\subsubsection{Hydrodynamic Trapping of Microswimmers}

We have shown that, for certain field strengths and directions $\mathcal{H},\phi_0$, swimmers can become collimated at a particular height $y^*$, thereby trapping them at a fixed vertical location. Here, we show that a swimmer's motility can additionally be harnessed to trap it in the horizontal direction, thereby localizing it to a point in space. To illustrate this effect, we consider motile HT-S swimmers, for which, when the field direction $\phi_0$ lies in the second and third quadrants, the swimmers focus at a height (see Eq. \eqref{eq:ydotFPB}) 
\begin{eqnarray}
y^{*}=\frac{R^2}{v_f}\frac{\mathcal{H}\sin{\phi_0}}{1-G},
\label{eq:ystar}
\end{eqnarray}
and attain an upstream orientation $\theta^*=\pi$, such that nearby trajectories converge to this height. Once motility is introduced, self-propulsion opposes the local flow while the swimmer remains oriented upstream, reducing its streamwise velocity. At sufficiently weak flow speeds, self-propulsion can exactly balance the flow advection, causing the swimmer to become stationary in both $x$ and $y$. The flow therefore acts as a \textit{hydrodynamic trap}, fixing the swimmer in space.

\smallskip

To estimate the flow speed required for trapping, we consider a spherical HT-S swimmer ($G=0$) and neglect the Fax\'en correction to the flow-advection speed as an approximation. The $\dot{x}$ equation \eqref{eq:xdotHTS_poiseuille}, with this approximation, then gives the trapping condition at the focusing height $y^*_B$ as $v_f(1-y_B^{*2}/R^2)\approx v^{\rm sp}$. Physically, this corresponds to a balance between the local Poiseuille flow speed and the upstream swimming speed. Substituting the expression for $y^*$ from Eq. \eqref{eq:ystar} into this balance condition yields the \textit{trapping flow} required to immobilize the swimmer:
\begin{eqnarray}
v_f^*=\frac{v^{\rm{sp}}+\sqrt{\left(v^{\rm{sp}}\right)^2 + \left(2R\mathcal{H}\sin{\phi_0}\right)^2}}{2}.
\label{expr:TrappingFlow}
\end{eqnarray}
For $\mathcal{H}=1\,\rm{s}^{-1}$ and $\phi_0=3\pi/4$, this gives $v_f^*\approx68\,\mu\text{m}/\text{s}$, trapping the swimmer at $y^*\approx R/2$. The resulting simulations of the hydrodynamic trapping effect for these parameters are shown in Fig. \ref{fig7}, in excellent agreement with our prediction.

\smallskip

Notably, if we have a dilute population of swimmers with multiple self-propulsion speeds, we can selectively trap swimmers with a particular $v^{\rm sp}$ from the population by adjusting the flow speed $v_f$ according to Eq. \eqref{expr:TrappingFlow}. The same principle can be extended to HT-A swimmers, providing a route to speed-selective trapping in motile shape-asymmetric populations as well.

\smallskip

\subsubsection{Shape-based Swimmer Sorting}\label{subsec:sorting}

\begin{figure}
\centerline{\includegraphics[width=\columnwidth]{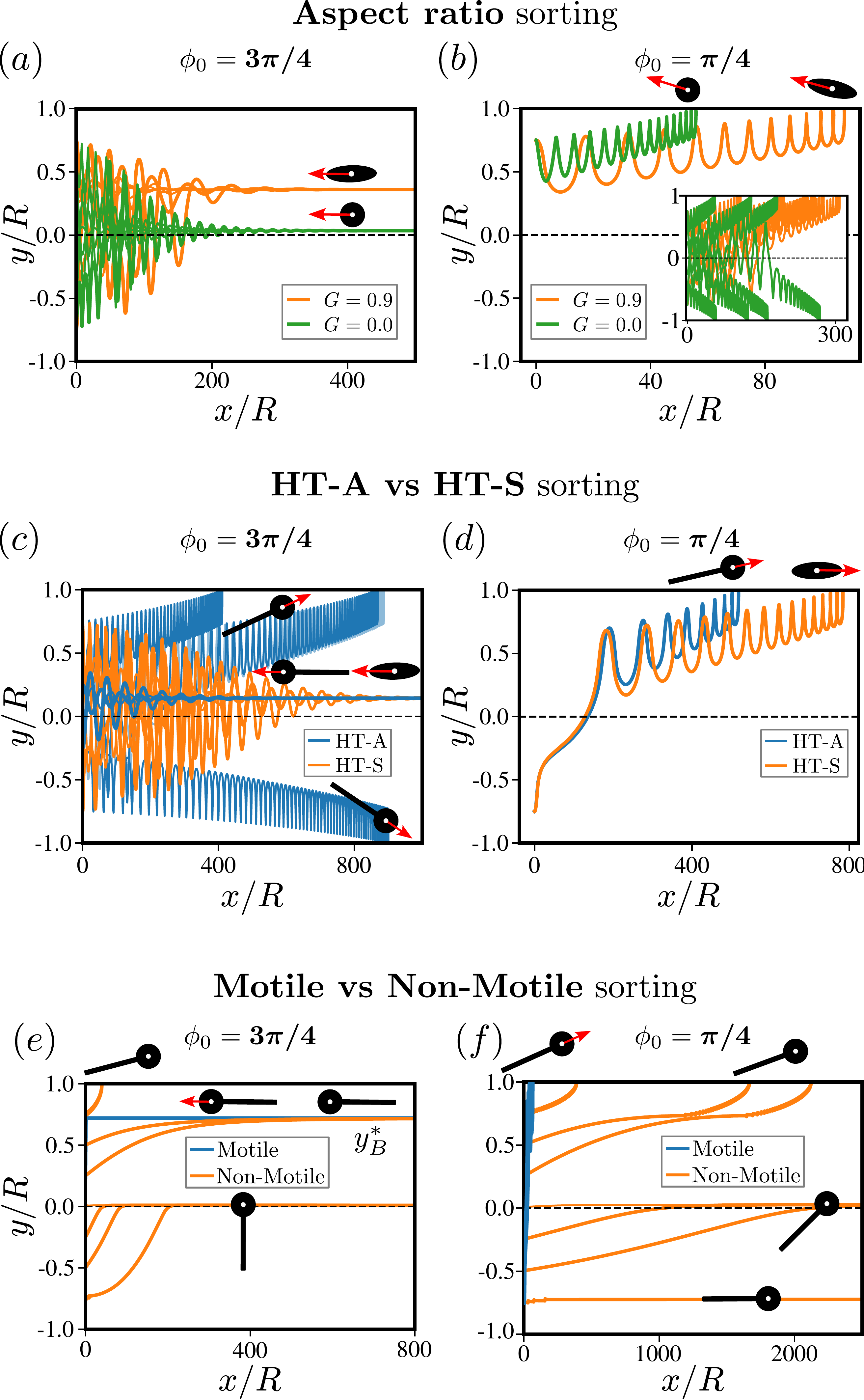}}
\caption{ \textbf{Sorting among Particle Classes} (a), (b) Sorting of HT-S particles by aspect ratio, for $\phi_0=3\pi/4$ and $\mathcal{H}=0.5\,\text{s}^{-1}$ in (a), and $\phi_0=\pi/4$ and $\mathcal{H}=0.5\,\text{s}^{-1}$ in (b). Inset of (b) shows trajectories for additional initial conditions. (c), (d) Sorting between HT-S and HT-A particles, for $\phi_0=3\pi/4$, $\mathcal{H}=0.2\,\text{s}^{-1}$ in (c), and $\phi_0=\pi/4$, $\mathcal{H}=0.5\,\text{s}^{-1}$ in (d). (e), (f) Sorting between motile and non-motile HT-A particles, for $\phi_0=3\pi/4$, $\mathcal{H}=1.0\,\text{s}^{-1}$ in (e), and $\phi_0=\pi/4$, $\mathcal{H}=1.0\,\text{s}^{-1}$ in (f).}

\label{fig8}
\end{figure}

The distinct transport responses of the four particle classes studied in this work, provide methods for sorting particles based on their shape as well as motility, which we now present below. 

\smallskip
\smallskip

\textbf{Aspect Ratio Sorting:}

\smallskip

\textbf{(a) Via collimation heights.} We first demonstrate how swimmers can be sorted based on differences in their aspect-ratio, through the resulting differences in their collimation heights [Fig. \ref{fig8}(a)]. With the field in the second quadrant, motile HT-S swimmers collimate at a unique height $y^*$. As shown in Fig. \ref{fig8}(a), spherical swimmers focus closer to the channel center, while progressively elongated swimmers focus at increasingly larger $y^*$. This shape dependence follows directly from the balance between flow- and field-induced rotation at the stationary orientation $\theta=\pi$, which determines $y^*$ through Eq. \eqref{eq:ydotFPB}. The aspect ratio enters through the coefficient $G$, which increases from $G=0$ for a sphere towards $G\to1$ for a highly elongated, needle-like particle. Equation \eqref{eq:ydotFPB} therefore predicts an increase of $y^*$ with $G$. Physically, elongation weakens the flow-induced rotation, requiring a stronger local shear, and hence a larger distance from the channel center, to balance the field-induced rotation. Therefore, swimmers with larger aspect ratios settle at larger collimation heights, providing a direct basis for aspect-ratio sorting.

\smallskip

\textbf{(b) Via wall focus-points.} A distinct form of aspect-ratio sorting emerges when the field lies in the first quadrant [Fig. \ref{fig8}(b)]. Here, swimmers continue to tumble because the field-induced rotation cannot balance the flow-induced rotation at the upstream orientation. The resulting anisotropic tumbling, coupled with self-propulsion $v^{\rm sp}$ and the HT-A contribution to cross-stream motion, $\dot{y}^{\rm HTA}_{\rm flow}$, produces shape-dependent wall-directed transport. Increasing the aspect ratio progressively biases swimmers towards the field-side wall: elongated swimmers preferentially reach this wall, but do so at progressively farther downstream focus-points, while spherical swimmers can reach either wall [inset of Fig. \ref{fig8}(b)]. This behavior arises from enhanced flow alignment with increasing elongation, which keeps the swimmer closer to the horizontal orientation and suppresses its cross-stream motion. This slows the approach to the field-side wall and increasingly suppresses trajectories towards the far-side wall. Thus, aspect ratio controls both the preferred wall and the location of its focus-point.

\smallskip
\smallskip
\smallskip
\smallskip

\textbf{Motile HT-S vs. HT-A Sorting:}

We first demonstrate sorting between motile HT-S and HT-A swimmers with the same Jeffery coefficient $G$, as shown in Fig. \ref{fig8}(c). For a field in the second quadrant, however, this matching is not essential: varying the aspect ratio of HT-S swimmers only shifts their focusing height and cannot drive them toward the walls. Thus, while HT-S swimmers focus at a unique height throughout the channel, HT-A swimmers are driven toward the walls in certain regions. Consequently, the swimmers reaching the walls are exclusively HT-A, providing spatial separation from their HT-S counterparts. 

\smallskip

In Fig. \ref{fig8}(d), we consider the same pair of swimmers, but with the field now in the first quadrant. Here, both HT-S and HT-A swimmers approach the walls, but their trajectories remain distinct: HT-A swimmers reach the walls earlier than HT-S swimmers. In this case, comparable aspect ratios are required to attribute the difference in arrival dynamics to the additional HT-A contribution to the cross-stream velocity, $\dot{y}^{\rm HTA}_{\rm flow}$ [see Eq. \eqref{eq:ydotHTA_poiseuille}]. Thus, even when both swimmer types ultimately reach the wall, their distinct wall focus-points provide a basis for sorting along the wall.

\smallskip
\smallskip

\textbf{Motile vs. Non-Motile Sorting:}

In Fig. \ref{fig8}(e), we sort motile and non-motile swimmers with a field at $\phi_0=3\pi/4$, and in Fig. \ref{fig8}(f) with a field at $\phi_0=\pi/4$. In both cases, $\mathcal{H}=1.0\,\text{s}^{-1}$.

\smallskip

For $\phi_0=3\pi/4$ [Fig. \ref{fig8}(e)], all motile swimmers approach the unique height $y^*_B$, whereas non-motile swimmers are focused at multiple heights, with some also approaching the walls. Thus, swimmers reaching the walls or collimating at heights other than $y^*_B$ are exclusively non-motile.

\smallskip

For $\phi_0=\pi/4$ [Fig. \ref{fig8}(f)], all motile swimmers approach the walls, whereas non-motile swimmers exhibit both wall-focusing and multi-height collimation. Thus, swimmers collimating within the channel are exclusively non-motile. Moreover, the wall focus-points of non-motile swimmers lie further downstream than those of their motile counterparts.

\smallskip

We have therefore demonstrated multiple ways to dynamically separate non-motile swimmers from a mixed-motility population, establishing a \textit{particle sieve} based on differences in motility and shape.

\subsubsection{Field and Flow Controls: Phase Diagrams}\label{subsec:PhaseDiagrams}

\begin{figure}
\centerline{\includegraphics[width=\columnwidth]{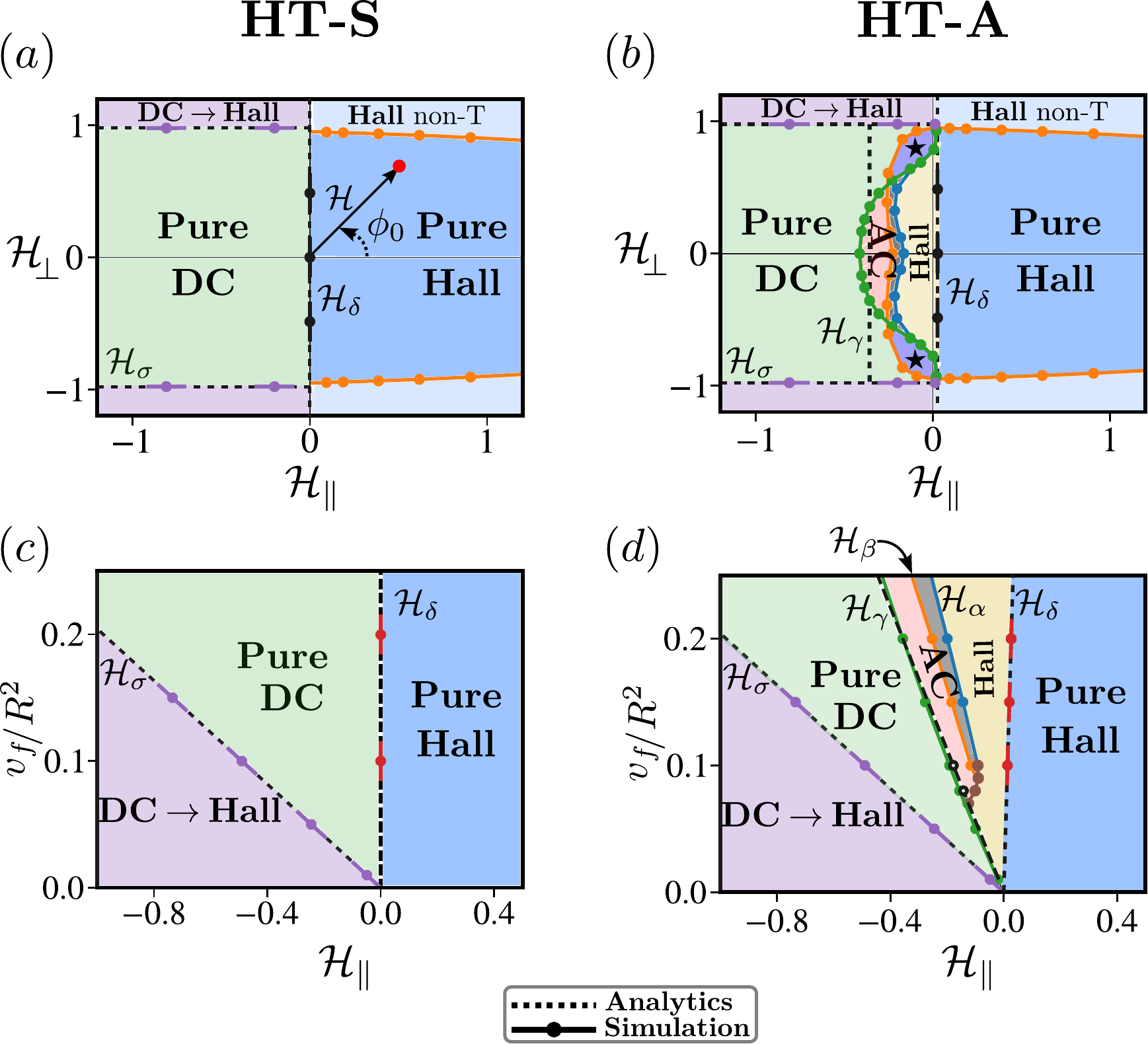}}
\caption{ \textbf{Phase diagrams} in $\mathcal{H}_{\parallel}$--$\mathcal{H}_{\perp}$ space for (a) HT-S and (b) HT-A swimmers. (c), (d) Corresponding phase diagrams in $v_f/R^2$--$\mathcal{H}_{\parallel}$ space for HT-S and HT-A swimmers, respectively. Here, $\mathcal{H}_{\parallel}=\mathcal{H}\cos\phi_0$ and $\mathcal{H}_{\perp}=\mathcal{H}\sin\phi_0$.}

\label{fig9}
\end{figure}

Our results for individual swimmer dynamics imply that, when a dilute suspension of motile swimmers is placed in the channel, three distinct population states emerge, shown schematically in Fig. \ref{fig1}(b): (i) \textbf{Hall} states, in which swimmers are driven towards and ultimately reach the channel walls; (ii) \textbf{DC} states, in which swimmers are guided towards and collimate at a fixed height; and (iii) \textbf{AC} states, in which swimmers collimate into persistent, constant-amplitude oscillatory motion.

To reveal how these states are selected by the external parameters, we construct comprehensive phase diagrams in Fig. \ref{fig9}, with each point labeled by the dominant state of the swimmer population, exposing the regions of parameter space in which each transport mode prevails.

\smallskip 

\textbf{Salient findings}: Head--tail shape asymmetry fundamentally reshapes the phase diagrams. While motile HT-S populations switch between pure Hall and DC states [Figs. \ref{fig9}(a),(c)], HT-A populations additionally access AC states and mixed regimes. These HT-A-specific regimes are most prevalent for fields oriented along the flow, $\mathcal{H}_{\perp}=\mathcal{H}\sin{\phi_0}=0$, as seen in the $\mathcal{H}_{\perp}$--$\mathcal{H}_{\parallel}$ phase space [Fig. \ref{fig9}(b)], and expand with increasing flow strength, as seen in the $v_f/R^2$--$\mathcal{H}_{\parallel}$ phase space [Fig. \ref{fig9}(d)], where $\mathcal{H}_{\parallel}=\mathcal{H}\cos{\phi_0}$. Sharp boundaries separate the resulting dynamical regimes, several of which we predict analytically, providing quantitative criteria for steering swimmer populations between distinct modes of transport through the imposed field and flow.

\smallskip

\textbf{Quantifying population states:} To determine the dominant state at each point in the phase diagrams, we use the fractional area of $y$-$\theta$ phase space approaching the dynamical attractor governing that state (see Ref. \cite{GomesAdhyapak_Field-Particle2026}). For example, the fractional area attracted to the unstable LC attractor in Fig. \ref{fig6}(b)-(ii) gives the fraction of swimmers ultimately reaching FP B, and hence the DC population state.

\smallskip

\textbf{Phase Boundaries:} Each of the phase diagrams contain sharp boundaries that mark qualitative changes in the underlying swimmer dynamics and can, in several cases, be traced to distinct physical mechanisms. The field $\mathcal{H}_{\delta}$ marks a stability-driven Hall--DC transition through the stability change of the relevant FP [Eq. \eqref{expr:Hdelta}], with excellent agreement between theory and simulations. The field $\mathcal{H}_{\sigma}$ instead marks a wall-induced DC–Hall transition, occurring when the collimation height is pushed to the channel wall, $y^*=R$. Using Eq. \eqref{eq:ydotFPB} with $y^*=R$ gives
\begin{eqnarray}
    \mathcal{H}_{\sigma}=\frac{v_f}{R}\frac{(1-G)}{\sin{\phi_0}}.
    \label{expr:Hsigma}
\end{eqnarray}
This prediction again agrees excellently with simulations, as seen from the $\mathcal{H}_{\sigma}$ lines in Figs. \ref{fig9}(c),(d). The $\mathcal{H}_{\gamma}$ boundary marks a nonlinear DC--AC transition and requires the dynamics of the stable LC, whose bifurcations are captured analytically in Ref. \cite{GomesAdhyapak_Field-Particle2026}. Using this theory, we obtain $\mathcal{H}_{\gamma}$ across the phase diagrams, with good agreement with simulations despite the presence of strong nonlinearities.

The remaining boundaries, $\mathcal{H}_{\alpha}$ and $\mathcal{H}_{\beta}$, involve stronger nonlinear effects and are therefore determined numerically; $\mathcal{H}_{\alpha}$ separates Hall and AC regimes, while $\mathcal{H}_{\beta}$ bounds two less prominent mixed-population regimes.

Together, these analytical and numerical boundaries expose how shape asymmetry reshapes the accessible population states, enable \textit{dynamical state switches} between distinct attractors by varying external field and flow parameters, and provide quantitative guidance for enhancing or suppressing HT-A-specific transport.

\section{Conclusions and Outlook}\label{sec:conclusion}

We have systematically investigated how shape asymmetry and motility, together with orienting fields and imposed flows, control microswimmer dynamics. We considered all four combinations of head--tail symmetry and motility, identified and characterized their dynamical attractors across field strengths and directions, and connected their phase-space structure to swimmer trajectories and applications.

\smallskip

For \textbf{non-motile HT-S particles}, we find distinct orientational dynamics in different channel layers under the same field, arising from the non-uniform shear of Poiseuille flow. We predict analogous orientational layering in other non-uniform shear flows. For \textbf{non-motile HT-A particles}, shape asymmetry additionally couples the flow to translation, producing layer-dependent cross-stream dynamics and multiple collimation heights; consequently, their transport can change substantially with the initial channel height. In the $\mathcal{H}\to0$ limit, which has been studied previously for HT-S particles and microswimmers \cite{JefferyPRSCA1922,salima_prl} and HT-A microswimmers \cite{GomesAdhyapakPRL2025}, we find that, for the hitherto overlooked non-motile HT-A case, particles exhibit oscillatory trajectories that follow streamlines on average, which may explain similar trajectories observed experimentally for immotile \textit{T. brucei} \cite{Stark_blood_flow}. Motility dissolves the horizontal layering in both particle classes. \textbf{Motile HT-S microswimmers} exhibit either wall-focused or collimated dynamics throughout the channel, with the trajectory class strongly selected by the field direction $\phi_0$. \textbf{Motile HT-A microswimmers}, in contrast, lose this strong field-quadrant selectivity. However, at weak to intermediate field strengths, wall-focused and collimated states can coexist for the same field parameters, together with a distinct stable-oscillatory state. Their trajectory state depends on both the initial height and orientation, while at strong fields their dynamics collapse onto those of HT-S swimmers.

\smallskip

A major theoretical advance of this work is the systematic identification of the dynamical attractors accessible to all four particle classes, their stability and variation with field strength and direction, and their correspondence with swimmer trajectories. More broadly, we show how the distinct HT-A flow-induced torques and force couple to fields and motility to generate qualitatively different dynamics. While the competition between HT-A torques and orienting fields was also established in our previous work \cite{GomesAdhyapak_Field-Particle2026}, here we show for the first time the consequences of the HT-A force generated by flow gradients and its competition with motility. When unmasked by activity, this force drives cross-stream motion even in non-motile particles, producing multiple collimation heights under a single field. The present study considers the limiting cases of strong activity, where motility masks the HT-A force, and zero activity, where the HT-A force acts without competition from motility; the regime in which the two effects become comparable and compete with each other remains an important direction for future study.

\smallskip

A major practical advance of our work is the set of application-oriented possibilities we have demonstrated. We demonstrate \textbf{hydrodynamic trapping}, in which motility can be exploited to localize a swimmer with a target self-propulsion speed to a point in the channel [Fig. \ref{fig7}], and show that differences in particle elongation, shape symmetry, and motility can be exploited for \textbf{particle sorting} [Fig. \ref{fig8}]. We further demonstrate \textbf{phase control}, showing how field strength and direction, flow speed, and channel width can drive individual and population-level transitions between distinct dynamical states through nonlinear transitions, as summarized by our phase diagrams for both HT-S and HT-A swimmers [Fig. \ref{fig9}]. Beyond these demonstrated applications, our analysis makes quantitative predictions, such as the saturation of the wall-focusing rate as $\mathcal{H}$ is varied  [Fig. \ref{fig6}(b)-(i)]. Beyond a certain field strength, the focusing rate becomes independent of $\mathcal{H}$ and is instead determined by the particle's self-propulsion speed.

\smallskip

Our framework extends naturally beyond the channel flows considered here to more general field--flow environments and provides a basis for continuum theories of shape-asymmetric constituents in fields and flows. It also offers a starting point for incorporating particle non-axisymmetry into field--flow coupled dynamics of shape-asymmetric microswimmers, opening new avenues for research in ``shape-asymmetric'' active matter.

\bmsubsection*{Acknowledgments}
TCA acknowledges grants CRG/2021/004759 from the
Science and Engineering Research Board (India) and MoE-STARS/STARS-2/2023-0814
for financial support. DCG acknowledges PMRF, Govt.  of India, for fellowship
and funding. TCA and DCG express gratitude to the \emph{Indian Institute of Science
Education and Research (IISER) Tirupati} for funds and facilities.
Furthermore, the support and the resources provided by `PARAM Brahma Facility'
under the National Supercomputing Mission, Government of India at the
\emph{Indian Institute of Science Education and Research (IISER) Pune} are
gratefully acknowledged.

\bmsubsection*{Conflicts of Interest}

The authors declare no conflicts of interest.

\bmsubsection*{Author Contributions}

DCG and TCA contributed equally to this work.

\bmsubsection*{Data Availability Statement}

The data that support the findings of this study are available from the corresponding author upon reasonable request.


\providecommand{\noopsort}[1]{}\providecommand{\singleletter}[1]{#1}%


\end{document}